# The Thermodynamic Limit of Indoor Photovoltaics Based on Energetically-Disordered Molecular Semiconductors

*Austin M. Kay[1], Maura E. Fitzsimons[1], Gregory Burwell[1], Paul Meredith[1], Ardalan Armin[1], and Oskar, J. Sandberg[1]*

[1]Sustainable Advanced Materials (Sêr-SAM), Centre for Integrative Semiconductor Materials (CISM), Department of Physics, Swansea University Bay Campus, Swansea SA1 8EN, United Kingdom

Email: ardalan.armin@swansea.ac.uk; o.j.sandberg@swansea.ac.uk

## Abstract

Due to their tailorable optical properties, organic semiconductors show considerable promise for use in indoor photovoltaics (IPVs), which present a sustainable route for powering ubiquitous "Internet-of-Things" devices in the coming decades. However, owing to their excitonic and energetically disordered nature, organic semiconductors generally display considerable sub-gap absorption and relatively large non-radiative losses in solar cells. To optimize organic semiconductor-based photovoltaics, it is therefore vital to understand how energetic disorder and non-radiative recombination limit the performance of these devices under indoor light sources. In this work, we explore how energetic disorder, sub-optical gap absorption, and non-radiative open-circuit voltage losses detrimentally affect the upper performance limits of organic semiconductor-based IPVs. Based on these considerations, we provide realistic upper estimates for the power conversion efficiency. The energetic disorder, inherently present in molecular semiconductors, is generally found to shift the optimal optical gap from 1.83 eV to ~1.9 eV for devices operating under LED spectra. Finally, we also describe a methodology (accompanied by a computational tool with a graphical user interface) for predicting IPV performance under arbitrary illumination conditions.

Using this methodology, we estimate the indoor PCEs of several photovoltaic materials, including the state-of-the-art systems PM6:Y6 and PM6:BTP-eC9.

## 1. Introduction

A recent coalescence of several technological trends has led to rapid developments in low-power networked devices, collectively referred to as the "Internet-of-Things" (IoT), which are poised to revolutionize almost all sectors of the global economy. [1, 2] While many of these devices may consume less than a microwatt of power, their aggregate energy consumption and environmental footprint must be carefully considered as they become ubiquitous in our homes and workspaces. [4-7] In this regard, indoor photovoltaics (IPVs) have emerged as a very attractive alternative for powering IoT devices. [8] The development of IPVs is propelled by progress in efficient charge controllers and supercapacitors, extending their viability for powering IoT devices to situations where illumination is not continuous. [9] Additionally, the lower light intensities and milder environments usually present indoors also provide less challenges for developing IPVs with enhanced longevities. [8, 10]

From a material optimization perspective, the criteria that IPVs are benchmarked against differ in several ways from those used for conventional photovoltaics. These differences stem from the fact that the emission spectra and irradiances of artificial sources, such as light-emitting diodes (LEDs), are quite unlike the standard AM1.5G spectrum of sunlight. In general, the spectral emission peaks of artificial light sources are narrower and centered at higher photon energies than the sunlight spectrum, and their integrated irradiances are usually at least three orders of magnitude lower. Because of these differing spectral characteristics, the optimal semiconductor energy gap needed for IPV applications is generally between around 1.7 eV and 1.9 eV, which is considerably wider than the bandgaps of conventional materials like crystalline silicon (1.1 eV), gallium arsenide (1.42 eV), and cadmium telluride (1.44 eV). [11]. For a given spectrum, the optimal gap and power conversion efficiency (PCE) is commonly estimated using the Shockley-Queisser (SQ) model. Under typical indoor conditions, the SQ model predicts PCEs surpassing 50% in single-junction devices with optimal energy gaps – significantly larger than the predicted PCE of 33.7% for AM1.5G sunlight. [12, 13] To achieve such theoretically-high PCEs, alternative wide-gap semiconductors are urgently needed for use in indoor applications. Next-generation, molecular semiconductors exhibit several attributes that make them desirable for such applications, including mechanical and form factor flexibility, low embodied energy manufacturing, and the fact that they are amenable to solution-based fabrication techniques like spin-coating and roll-to-roll printing. [14, 15] Of these, organic semiconductors are of particular technological-relevance for indoor applications because of the vast palette of materials available and the tunability afforded by synthetic organic chemistry. [16-19]

In recent years, the performance of organic photovoltaics (OPVs) based on combinations of polymeric donors and low-offset, non-fullerene (small molecule) acceptors (NFAs) has advanced considerably. [20-25] OPV materials and device architectures, however, are not yet optimized for indoor applications, due in part to the relative infancy of the field and the lack of established measurement standards. [26-28] Furthermore, the maximum, experimentally-determined PCE reported for an IPV device based upon conventional OPV principles is currently around 31%, whereas typical PCEs are on the order of 20% – considerably lower than the thermodynamic limit calculated via the SQ model. [29, 30] A thorough investigation of the realistic thermodynamic limits of existing OPVs for indoor applications is therefore required for two reasons. Firstly, such an investigation would provide a roadmap for next-generation IPV development, including which routes for device optimization should be pursued. Secondly, it would provide a benchmark for IPV device characterization – until relevant standards are established, inter-laboratory comparisons are complicated by sources of uncertainty and error. [26] These include variations in the spectra and irradiances used to simulate indoor illuminations – all too common problems encountered in the early days of organic solar cells designed for outdoor power generation, but re-emerging now for IPVs.

To obtain realistic predictions for the maximum PCEs and optimal gaps of IPVs based on organic semiconductors, the associated loss mechanisms of OPV devices must be audited. [9] This includes accounting for the excitonic nature of OPVs, as well as the associated static disorder that correlates with a broadened absorption onset and increased sub-gap absorption. [31-33] In general, absorption well below the optical gap induces radiative losses in the open-circuit voltage. Sub-gap absorption is typically correlated with a so-called Urbach energy ($E_{\mathrm{U}}$) – a measure of the exponential decay in absorption with decreasing photon energy ($E$). [34, 35] As a result, OPVs with lower energetic disorder and smaller $E_{\mathrm{U}}$ are likely to have reduced open-circuit voltage losses and, consequently, higher PCEs. [22, 36-38] In addition to the radiative open-circuit voltage losses induced by sub-gap absorption, further non-radiative open-circuit voltage losses are present in OPVs due to the intrinsic prevalence of non-radiative recombination. [39-41] The electroluminescent external quantum efficiency ($\mathrm{EQE_{EL}}$) is commonly used to estimate these non-radiative open-circuit voltage losses. [42] While numerous processes can contribute to the non-radiative recombination in OPVs, the associated voltage loss has been found to generally correlate with the energy gap. [39, 41]

In this work, we step beyond the rudimentary SQ model to make realistic predictions for the PCEs of existing OPVs in indoor settings. We explore the effects of the optical gap and energetic disorder on the optimal PCE. In addition, we investigate the role of non-radiative open-circuit voltage losses, while accounting for the energy gap-dependence of non-radiative recombination using an optimistic-yet-realistic empirical model guided by literature OPV data. Following this, we present a methodology and an accompanying computational tool (with an accessible graphical user interface) for recontextualizing a given photovoltaic system's existing measurements under one-Sun conditions to predict how it might perform under arbitrary illumination conditions. Utilizing this methodology, which employs measurements of a device's photovoltaic external quantum efficiency spectrum and its open-circuit voltage under AM1.5G conditions, we predict the indoor performance of dozens of emerging OPV systems. Finally, we demonstrate that the "fruit fly" systems PM6:Y6 and PM6:BTP-eC9 are likely limited to PCEs below 20% in indoor settings.

## 2. Results and Discussion

## 2.1. Photovoltaic Figures-of-Merit

The spectral fingerprints of artificial sources of light generally differ from source to source, displaying variations in intensity and separation of emission peaks. In **Figure 1a**, the spectral photon flux densities ($\Phi_{\mathrm{source}}$) of typical indoor light sources, including the 'warm white' 2700K LED and 'cool white' 4000K LED, are illustrated alongside the International Commission on Illumination's (CIE's) standard illuminant LED-B4. Therein, the integrated power density of each source, $P_{\mathrm{source}} = \int_0^{\infty} E\, \Phi_{\mathrm{source}}(E)\, \mathrm{d}E$, is scaled to a total illuminance of $500$ lux (see **Section S1** of the **Supporting Information**). The corresponding scaled AM1.5G spectrum has been included for comparison. In this work, we primarily consider the CIE LED-B4 standard as the indoor light source as LEDs are becoming more commonplace in most indoor settings. However, it should be noted that the obtained findings are largely independent of the used LED source; similar results are found for the 2700K LED and 4000K LED spectra (see **Supporting Information**). Additional discussions for other standard indoor light sources, including the CIE FL-2, CIE FL-7, and CIE FL-11 spectra, are also available in the **Supporting Information**. [43, 44]

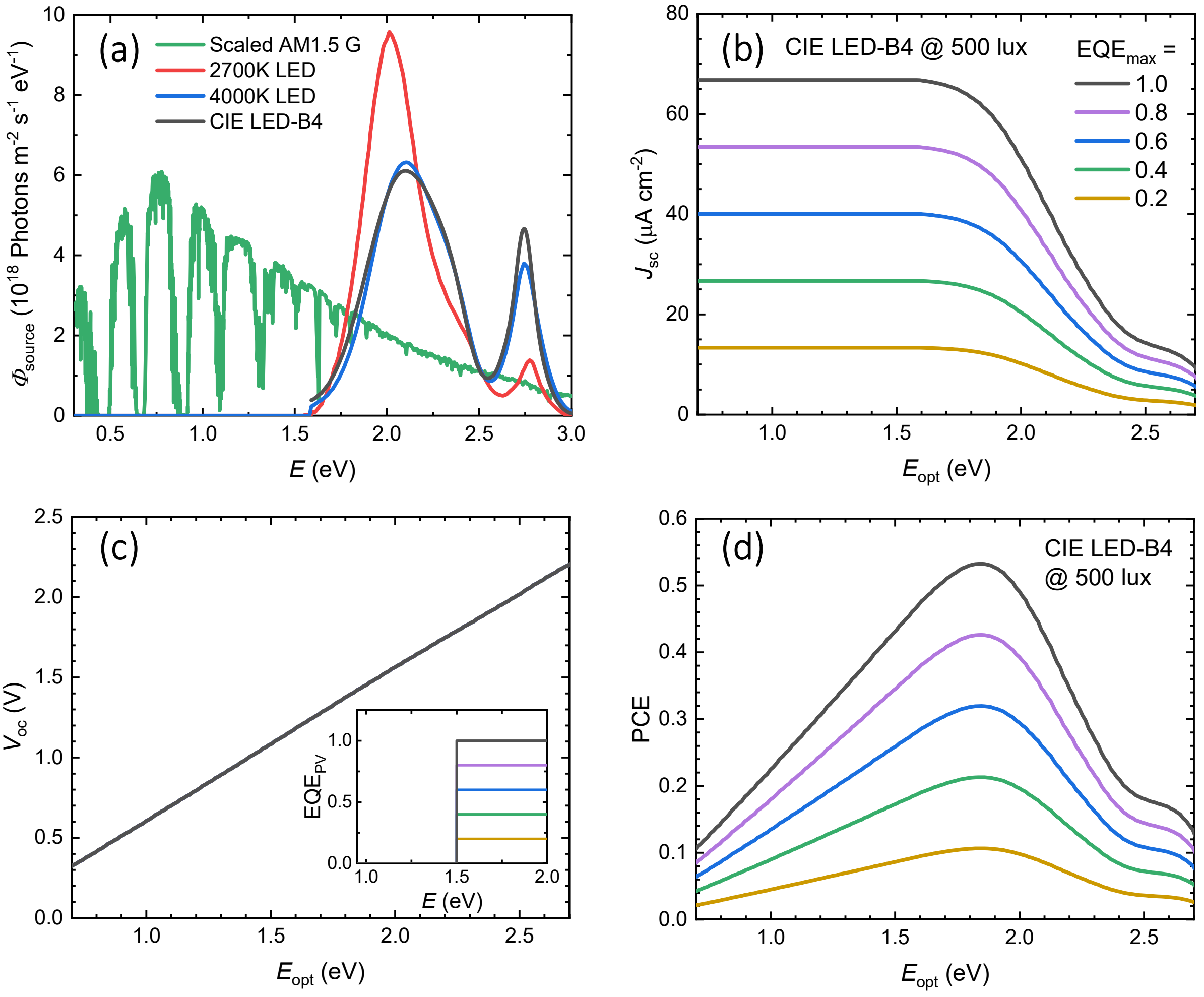

**Figure 1**: **(a)** The scaled AM1.5G spectrum for sunlight (green), the 2700K LED spectrum (red), the 4000K LED spectrum (blue), and the CIE LED-B4 spectrum (black), all plotted against the photon energy at an illuminance of 500 lux. In **(b)**, **(c)**, and **(d)**, the short-circuit current density, the open-circuit voltage, and the power conversion efficiency under the LED-B4 spectrum at 500 lux, respectively, are plotted against the optical gap for varied $\mathrm{EQE}_{\mathrm{max}}$ , assuming the step-function model for $\mathrm{EQE}_{\mathrm{PV}}$ given by Equation **(7)**. The effect of increasing $\mathrm{EQE}_{\mathrm{max}}$ is illustrated for an optical gap $E_{\mathrm{opt}} = 1.5$ eV in the inset graph in **(c)**.

Under illumination, a photovoltaic device will generate power at efficiency [11]

$$\mathrm{PCE} = \frac{\mathrm{FF}\, J_{\mathrm{sc}}\, V_{\mathrm{oc}}}{P_{\mathrm{source}}}. \quad \textbf{(1)}$$

Here, $V_{\mathrm{oc}}$ is the open-circuit voltage, $J_{\mathrm{sc}}$ is the short-circuit current density, and FF is the fill factor. In general, the open-circuit voltage and the short-circuit current density relate to the device's photovoltaic external quantum efficiency $\mathrm{EQE_{PV}}(E)$ (the ratio of the number of collected charge carriers to the number of incident photons at a given photon energy $E$) via [11]

$$V_{\mathrm{oc}} = \frac{kT}{q} \ln\left[1 + \frac{J_{\mathrm{sc}}}{J_0(V)}\right], \quad \textbf{(2)}$$

$$J_{\mathrm{sc}} = q \int_0^{\infty} \mathrm{EQE_{PV}}(E)\ \Phi_{\mathrm{source}}(E)\, \mathrm{d}E. \quad \textbf{(3)}$$

Wherein $k$ denotes the Boltzmann constant, $q$ the elementary charge, and $T$ the temperature. The quantity $J_0$, on the other hand, is the dark saturation current density. It is calculated using $J_0(V) = J_0^{\mathrm{rad}}(V)/\mathrm{EQE_{EL}}$, where the radiative dark saturation current density ($J_0^{\mathrm{rad}}$) is defined by [42]

$$J_0^{\mathrm{rad}}(V) = q \int_0^{\infty} \mathrm{EQE_{PV}}(E) \Phi_{\mathrm{bb}}(E) w(E,V)\, \mathrm{d}E, \quad \textbf{(4)}$$

where $\Phi_{\mathrm{bb}}(E) = \frac{2\pi E^2}{h^3 c^2} \exp\left(-\frac{E}{kT}\right)$ is the spectral photon flux density of the ambient black-body radiation at thermal equilibrium ($h$ is the Planck constant and $c$ is the speed of light). Here, $w(E,V)$ is a degeneracy factor accounting for state-filling effects; [36] in the thermodynamic limit, $w(E,V)$ can be approximated as (see **Section S3** of the **Supporting Information**):

$$w(E,V) = \frac{1}{\left[1 + \exp\left(\frac{qV - E}{2kT}\right)\right]^2}. \quad \textbf{(5)}$$

For above-gap states ($E \gg qV$) in the non-degenerate limit, $w(E,V) = 1$ typically applies and $J_0^{\mathrm{rad}}$ is independent of the voltage. For the general case, however, $J_0^{\mathrm{rad}}$ depends on the voltage and an iterative approach is used to evaluate Equation **(2)** (again, see **Section S3** of the **Supporting Information**). Note that the device's $\mathrm{EQE_{EL}}$ equals one in the radiative limit, giving $J_0 = J_0^{\mathrm{rad}}$ while $V_{\mathrm{oc}} = V_{\mathrm{oc}}^{\mathrm{rad}}$ ($V_{\mathrm{oc}}^{\mathrm{rad}}$ is the corresponding radiative $V_{\mathrm{oc}}$). Photovoltaic devices are generally far from the radiative limit; non-radiative recombination increases $J_0$ which, in turn, reduces the open-circuit voltage as $V_{\mathrm{oc}} = V_{\mathrm{oc}}^{\mathrm{rad}} - \Delta V_{\mathrm{oc}}^{\mathrm{nr}}$, where $\Delta V_{oc}^{nr}$ is the associated non-radiative open-circuit voltage loss given by $\Delta V_{oc}^{nr} = -\frac{kT}{q}\ln(\mathrm{EQE_{EL}})$ for $V_{oc} \gg kT/q$.[42]

Finally, we assume that the current density is approximated by $J = -J_{sc} + J_0^{rad}(V)\exp\left(\frac{q\Delta V_{oc}^{nr}}{kT}\right)\left[\exp\left(\frac{qV}{kT}\right) - 1\right]$, where $J_{sc}$ is given by Equation (3) and $J_0^{rad}(V)$ by Equation (4), corresponding to the case of ideal charge collection. Subsequently, the FF and the PCE are determined numerically using the iterative approach outlined in **Section S3** of the **Supporting Information**. However, we note that for $w = 1$ (and assuming $V_{oc}$ larger than 0.5 V), the fill factor is well-approximated by[45]

$$\mathrm{FF} \approx \frac{\frac{qV_{oc}}{kT} - \ln\left(1 + \frac{qV_{oc}}{kT}\right)}{1 + \frac{qV_{oc}}{kT}}. \quad \textbf{(6)}$$

This suggests that, in the case of ideal charge transport, the leading-order behavior of the fill factor is primarily determined by the open-circuit voltage. Consequently, minimizing open-circuit voltage losses is of paramount importance for realizing high-PCE IPVs based on organic semiconductors. We note, however, that in reality the FF is influenced further by several additional factors; most notably the shunt resistance plays a crucial role in limiting the FF under low light indoor conditions.[8]

## 2.2. Effect of Radiative Open-Circuit Voltage Losses

We now consider the influence of radiative open-circuit voltage losses on the performance of IPVs by first discussing the idealized case of a sharp optical gap and no sub-gap absorption. In this case, $\mathrm{EQE_{PV}}$ can be modelled using a step function, where all photons of energy greater than or equal to a threshold optical gap ($E_{opt}$) generate a collected electron-hole pair at efficiency $\mathrm{EQE_{max}}$, whereas photons of energy less than the optical gap do not:

$$\mathrm{EQE_{PV}}(E) = \begin{cases} \mathrm{EQE_{max}}, & \text{if} \quad E \geq E_{opt}, \\ 0, & \text{otherwise.} \end{cases} \quad \textbf{(7)}$$

The photovoltaic external quantum efficiency in the SQ model is defined using this equation in the ideal case that $\mathrm{EQE_{max}} = 1$.[13] For an $\mathrm{EQE_{PV}}$ spectrum modeled using Equation **(7)**, the short-circuit current density, radiative open-circuit voltage, and resultant PCE under the CIE LED-B4 spectrum at 500 lux are shown for varying $\mathrm{EQE_{max}}$ in **Figure 1b, c,** and **d**, respectively. As shown, at a particular optical gap, the short-circuit current density is directly proportional to $\mathrm{EQE_{max}}$. The open-circuit voltage, however, is independent of $\mathrm{EQE_{max}}$ and so the $V_{\mathrm{oc}}$ curves are perfectly aligned and equal to the open-circuit voltage predicted by the SQ model ($V_{\mathrm{oc}}^{\mathrm{SQ}}$), which can be approximated as [45]

$$qV_{\mathrm{oc}}^{\mathrm{SQ}} \approx E_{\mathrm{opt}} - kT \ln\left[\frac{2\pi q}{h^3 c^2} \frac{E_{\mathrm{opt}}^2 kT}{J_{\mathrm{sc}}^{\mathrm{SQ}}}\right] \qquad \textbf{(8)}$$

for $qV_{\mathrm{oc}}^{\mathrm{SQ}} \ll E_{\mathrm{opt}}$, where $J_{\mathrm{sc}}^{\mathrm{SQ}} = q\int_{E_{\mathrm{opt}}}^{\infty} \Phi_{\mathrm{source}}(E)\,\mathrm{d}E$. Note that since the FF is determined by the $V_{\mathrm{oc}}$ in this case, the PCE predominantly scales in a similar way to $J_{\mathrm{sc}}$, changing linearly with $\mathrm{EQE_{max}}$.

From **Figure 1d**, it is evident that in the SQ model, the maximum PCE under the CIE LED-B4 spectrum at 500 lux is 53%, obtained at an optical gap $E_{\mathrm{opt}} = 1.83$ eV, with $V_{\mathrm{oc}}^{\mathrm{SQ}} = 1.41$ V and $J_{\mathrm{sc}}^{\mathrm{SQ}} = 62.1\ \mu\mathrm{A\ cm^{-2}}$. However, for current state-of-the-art OPVs, the empirical upper limit of the $\mathrm{EQE_{PV}}$ is closer to 0.85. Therefore, to realistically estimate the PCEs of IPVs based on organic semiconductors, an above-gap photovoltaic quantum efficiency of $\mathrm{EQE_{max}} = 0.85$ is herein assumed – unless explicitly stated otherwise – as this value describes realistically-high performance. The corresponding maximum PCE for $\mathrm{EQE_{max}} = 0.85$ is reduced to 45.3%, which is still obtained at $E_{\mathrm{opt}} = 1.83$ eV (for CIE LED-B4 at 500 lux).

Despite being rudimentary, the step-function model given by Equation **(7)** is a good approximation for $\mathrm{EQE_{PV}}$ in semiconductors with well-defined band edges, such as crystalline, inorganic semiconductors. Many photovoltaic materials, however, are not well-described by the highly-idealized step-function model. A more realistic prediction for the PCEs of IPVs based on energetically-disordered materials, including OPVs, must account for the inherent, static energetic disorder associated with the density of states. As increased static energetic disorder broadens the effective band edges and leads to increased sub-gap absorption, it will increase radiative open-circuit voltage losses and reduce the PCE.

Sub-gap absorption in disordered materials is commonly described by a tail that decays exponentially with decreasing photon energy below the gap. This tail may be designated a characteristic energy – the aforementioned Urbach energy ($E_{\mathrm{U}}$). [35] Consequently, a more realistic model for $\mathrm{EQE}_{\mathrm{PV}}$ in many photovoltaics is given by

$$\mathrm{EQE}_{\mathrm{PV}}(E) = \mathrm{EQE}_{\mathrm{max}} \begin{cases} 1, & \text{if} \quad E \geq E_{\mathrm{opt}}, \\ \exp\left(\dfrac{E - E_{\mathrm{opt}}}{E_{\mathrm{U}}}\right), & \text{otherwise.} \end{cases} \tag{9}$$

The Urbach energy correlates with the level of disorder in a system and, as illustrated in **Figure 2a**, it determines the gradient of the exponential decay of the sub-gap tail. A reasonable minimum value for the Urbach energy of OPVs is the thermal energy ($kT$); throughout the remainder of this work we assume $kT = 25.3\ \mathrm{meV}$ (corresponding to $T = 20°\mathrm{C} = 293.15\ \mathrm{K}$). [31]

The presence of sub-gap Urbach tails gives rise to a decrease in $V_{\mathrm{oc}}^{\mathrm{rad}}$, as shown in **Figure 2b**. In **Figure 2c**, it is shown that these losses, in turn, reduce the maximum power conversion efficiency from 45% to around 33% (in the $E_{\mathrm{U}} = 50\ \mathrm{meV}$ case), while concurrently blue-shifting the best-performing $E_{\mathrm{opt}}$ from 1.83 eV to 1.91 eV. Material systems with high $E_{\mathrm{U}}$ therefore require larger optical gaps to achieve high performance. We note that the short-circuit current density is found to be largely independent of $E_{\mathrm{U}}$. The loss in PCE shown in Figure 2c is therefore a result of the radiative open-circuit voltage loss ($\Delta V_{\mathrm{oc,sub-gap}}^{\mathrm{rad}}$) induced by sup-gap tails. This voltage loss is quantified by the deviation between $V_{\mathrm{oc}}^{\mathrm{SQ}}$ (determined in the SQ model) and the $V_{\mathrm{oc}}^{\mathrm{rad}}$ obtained in case of a sub-gap tail, $\Delta V_{\mathrm{oc,sub-gap}}^{\mathrm{rad}} = V_{\mathrm{oc}}^{\mathrm{SQ}} - V_{\mathrm{oc}}^{\mathrm{rad}}$. For the open-circuit voltage curves of Figure 2b, these deviations were determined then plotted in **Figure 2d**. For $E_{\mathrm{U}} \geq kT$, the optical gap-dependent behavior of these curves can be described by the following analytical approximations (see **Supporting Information**):

$$q\Delta V_{\mathrm{oc,sub-gap}}^{\mathrm{rad}} \approx \begin{cases} \left(\frac{E_{\mathrm{U}}}{kT}-1\right)\left(E_{\mathrm{opt}}-qV_{\mathrm{oc}}^{\mathrm{SQ}}\right)+E_{\mathrm{U}}\ln\left[\frac{\left(\frac{qV_{\mathrm{oc}}}{E_{\mathrm{opt}}}\right)^{2}}{1-\frac{kT}{E_{\mathrm{U}}}}\right], & \text{if } E_{\mathrm{U}}>kT. \\ kT\ln\left[\frac{E_{\mathrm{opt}}}{3kT}+1\right]+kT\ln\left[1-\left(\frac{qV_{\mathrm{oc}}}{E_{\mathrm{opt}}}\right)^{3}\right], & \text{if } E_{\mathrm{U}}=kT. \end{cases} \tag{10}$$

Equation **(10)** describes the behavior of $\Delta V_{\mathrm{oc,sub-gap}}^{\mathrm{rad}}$ for $E_{\mathrm{U}} \geq kT$ at typical optical gaps, as shown by the dashed curves in Figure 2d. We note that, in accordance with Equation **(10)**, for $E_{\mathrm{U}} > kT$ the associated radiative open-circuit voltage displays a $V_{\mathrm{oc}}^{\mathrm{rad}} \propto \frac{E_U}{kT} V_{\mathrm{oc}}^{\mathrm{SQ}}$ type dependence. This translates to a radiative ideality factor above unity, consistent with previous reports. [46, 47]

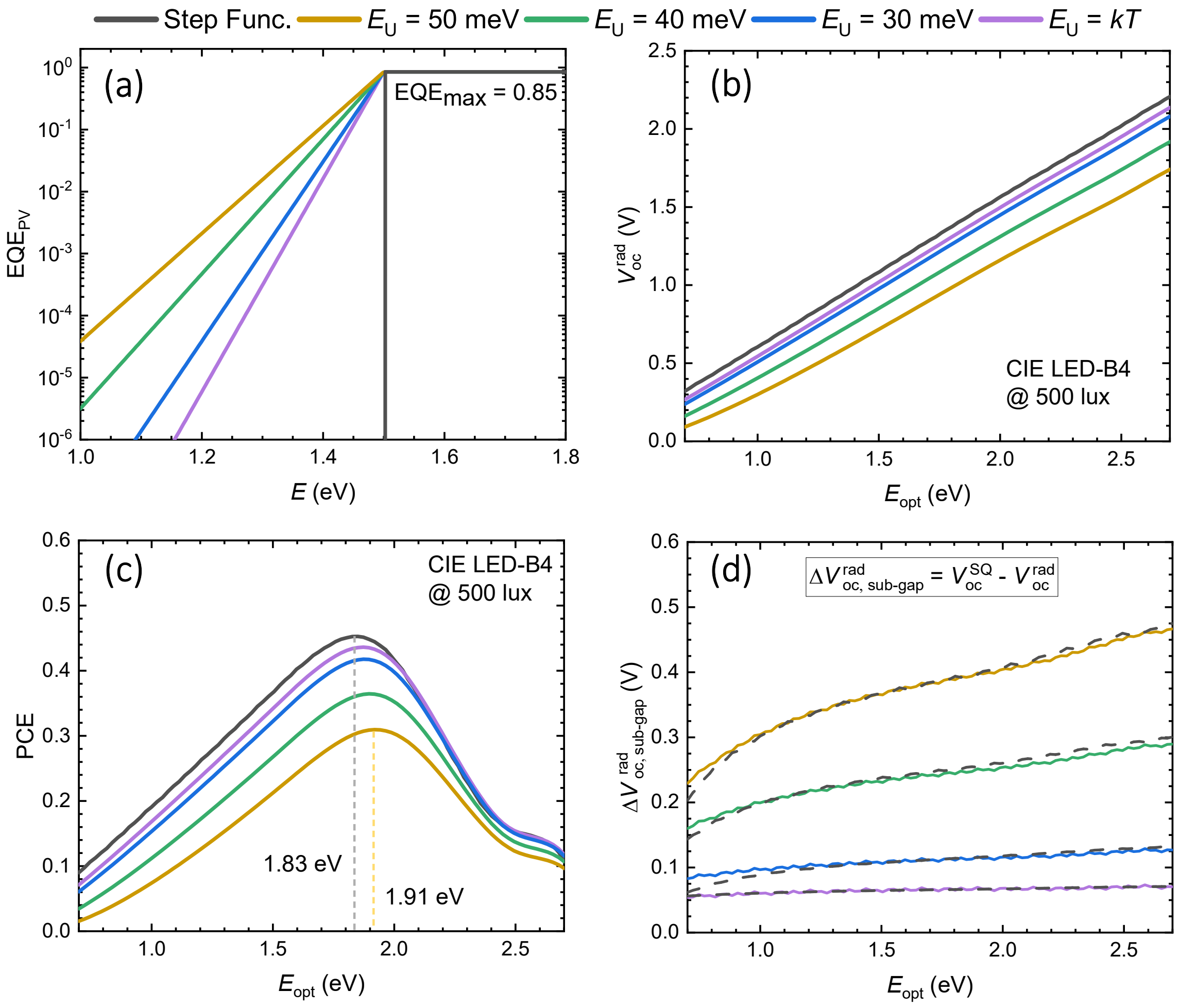

**Figure 2**: Investigating the effect of sub-gap tails of varying Urbach energy on the open-circuit voltage and the power conversion efficiency. **(a)** Photovoltaic external quantum efficiency spectra centered at an optical gap $E_{\mathrm{opt}} = 1.5$ eV, with $\mathrm{EQE}_{\mathrm{max}} = 0.85$ and $E_{\mathrm{U}}$ varied from 0 (step function) to 50 meV. **(b)** The resultant open-circuit voltages in the radiative limit, plotted as a function of the optical gap. **(c)** The PCE under the CIE LED-B4 spectrum at 500 lux, plotted as a function of the optical gap for a variety of Urbach energies. **(d)** The solid curves indicate the numerically-calculated deviations between the open-circuit voltage in the SQ model and the sub-gap Urbach tail model. The dashed lines indicate the corresponding analytical approximation given by Equation **(10)**.

Previously, the static energetic disorder in organic semiconductors has instead been frequently modelled in terms of a Gaussian distribution of states. Consistent with this, the $\mathrm{EQE}_{\mathrm{PV}}$ associated with excitonic sub-gap absorption in several low-offset NFA OPV material systems was recently found to be well-described by [31, 33]

$$\mathrm{EQE}_{\mathrm{PV}}(E) = \frac{\mathrm{EQE}_{\mathrm{max}}}{2}\left\{\exp\left[\frac{E - E_{\mathrm{opt}} + \frac{\sigma_{\mathrm{s}}^2}{2kT}}{kT}\right]\mathrm{erfc}\left[\frac{E - E_{\mathrm{opt}} + \frac{\sigma_{\mathrm{s}}^2}{kT}}{\sigma_{\mathrm{s}}\sqrt{2}}\right] + \mathrm{erfc}\left[\frac{E_{\mathrm{opt}} - E}{\sigma_{\mathrm{s}}\sqrt{2}}\right]\right\}, \quad \mathbf{(11)}$$

where $E_{\mathrm{opt}}$ is the centre of a Gaussian distribution of exciton states with static disorder parameter $\sigma_{\mathrm{s}}$. Here, $\mathrm{erfc}(x)$ denotes the complementary error function. The spectral behavior of Equation **(11)** at different $\sigma_{\mathrm{s}}$ is illustrated in **Figure 3a** for $\mathrm{EQE}_{\mathrm{max}} = 0.85$ and an optical gap of 1.5 eV. For energies well below the gap ($E \ll E_{\mathrm{opt}}$), Equation **(11)** reduces to a sub-gap Urbach tail with $E_{\mathrm{U}} = kT$. Above the gap, on the other hand, a saturation is reached wherein $\mathrm{EQE}_{\mathrm{PV}}(E) \rightarrow \mathrm{EQE}_{\mathrm{max}}$. Between these two regimes lies a transition regime with a shape and spectral broadness determined by $\sigma_{\mathrm{s}}$.

**Figure 3b** and **3c** show the $V_{\mathrm{oc}}^{\mathrm{rad}}$ and PCE as a function optical gap obtained based on the $\mathrm{EQE}_{\mathrm{PV}}$ (Equation **(11)**) from Figure 3a. The corresponding radiative open-circuit voltage losses $\Delta V_{\mathrm{oc,sub-gap}}^{\mathrm{rad}}$, induced by the sub-gap $\mathrm{EQE}_{\mathrm{PV}}$, are shown in **Figure 3d**. As illustrated throughout **Figure 3**, a higher static energetic disorder gives rise to increased radiative open-circuit voltage loss, thereby reducing the power conversion efficiency from 45% in the step function model to 37% in the $\sigma_s = 100$ meV case. In addition, the best-performing optical gap is once again blue-shifted from 1.83 eV to 1.88 eV in this case.

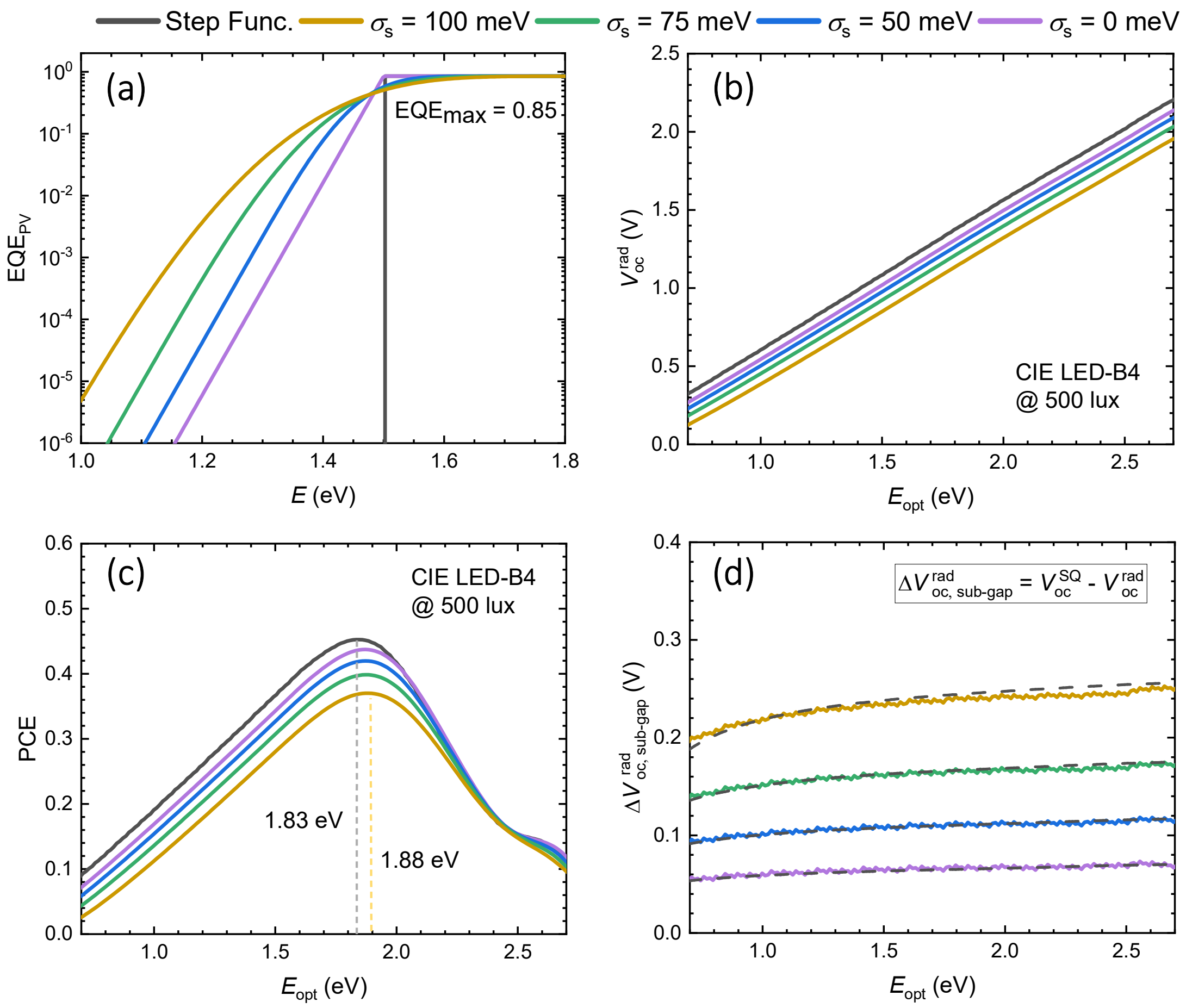

**Figure 3**: Investigating the effect of energetic disorder on the open-circuit voltage and the power conversion efficiency. **(a)** Photovoltaic external quantum efficiency spectra centered at an optical gap $E_{\mathrm{opt}} = 1.5$ eV, with $\mathrm{EQE}_{\mathrm{max}} = 0.85$ and $\sigma_s$ varied from 0 meV to 100 meV, plotted alongside the step function model for $\mathrm{EQE}_{\mathrm{PV}}$ (in black). **(b)** The resultant open-circuit voltages in the radiative limit, plotted as a function of the optical gap. **(c)** The PCE under the CIE LED-B4 spectrum at 500 lux, plotted as a function of the optical gap for varying $\sigma_s$. **(d)** The solid curves indicate the numerically-calculated deviations between the open-

circuit voltage in the SQ model and the OPV model, where $\mathrm{EQE_{PV}}$ is modelled in the latter using Equation **(11)**. The dashed lines indicate the analytical approximation given by Equation **(12)**.

As with the varied Urbach energy case, an analytical approximation for $\Delta V_{\mathrm{oc,sub-gap}}^{\mathrm{rad}}$ for the case of a sub-gap $\mathrm{EQE_{PV}}$ given by Equation **(11)** can be obtained assuming that the short-circuit current density is invariant of $\sigma_{\mathrm{s}}$ and solely determined by the contribution from the above-gap $\mathrm{EQE_{PV}}$ (see **Section S4** of the **Supporting Information**). Under such conditions, $\Delta V_{\mathrm{oc,sub-gap}}^{\mathrm{rad}}$ can be obtained from

$$q\Delta V_{\mathrm{oc,sub-gap}}^{\mathrm{rad}} \approx \frac{\sigma_{\mathrm{s}}^2}{2kT} + kT\ln\left[\frac{E_{\mathrm{opt}}}{3kT}\left(1-\frac{\sigma_{\mathrm{s}}^2}{E_{\mathrm{opt}}kT}\right)^3\right] + kT\ln\left[1-\left(\frac{qV_{\mathrm{oc}}}{E_{\mathrm{opt}}}\right)^3\right]. \quad \textbf{(12)}$$

where the last term on the right-hand-side is a correction accounting for state filling effects ($w \neq 1$). Equation **(12)** is indicated by the dashed lines in **Figure 3d**. As shown, the approximation agrees well with the numerically-calculated results for typical optical gaps.

## 2.3. Effect of Non-Radiative Open-Circuit Voltage Losses

In real photovoltaic devices, the open-circuit voltage is further reduced by non-radiative recombination, which reduces $\mathrm{EQE_{EL}}$ below unity and gives rise to a non-zero non-radiative open-circuit voltage loss $\Delta V_{\mathrm{oc}}^{\mathrm{nr}}$. [39, 42, 48] In OPVs, the non-radiative open-circuit voltage loss measured under one Sun has been observed to increase with decreasing energy gap, consistent with the energy-gap law. [39, 41, 42, 48] This is demonstrated in **Figure 4a**, where experimental $\Delta V_{\mathrm{oc}}^{\mathrm{nr}}$ data compiled by Ullbrich et al. [39] are plotted against the energy of the CT state ($E_{\mathrm{CT}}$), which we take as a proxy for $E_{\mathrm{opt}}$ (valid for low-offset, NFA OPV blends). Additional $\Delta V_{\mathrm{oc}}^{\mathrm{nr}}$ data for systems with fullerene acceptors and NFAs are plotted as blue squares and green triangles, respectively. We note that at light intensities representative of indoor settings, the non-radiative loss may, in general, be larger (due to additional trap-assisted recombination); as such, the data in **Figure 4a** could be considered as a realistic upper estimate of $\Delta V_{\mathrm{oc}}^{\mathrm{nr}}$ in organic semiconductor-based IPVs.

To obtain a realistic estimate of non-radiative open-circuit voltage losses in state-of-the-art OPVs, we have designed an empirical, qualitative model for $\Delta V_{oc}^{nr}$ based on the experimental data in **Figure 4a**. In this empirical model, $\Delta V_{oc}^{nr}$ is modelled as a quadratic of the form

$$\Delta V_{oc}^{nr} = \begin{cases} AE_{opt}^2 + BE_{opt} + C, & \text{if} \quad E_{opt} \leq 2.601 \text{ eV}, \\ 0.0945 \text{ V}, & \text{otherwise}, \end{cases} \quad \textbf{(13)}$$

where the optical gap has units of eV, and the coefficients are $A = 0.123 \text{ V}/(\text{eV})^2$, $B = -0.64 \text{ V}/(\text{eV})$, and $C = 0.927$ V. The transition at $2.601$ eV prevents $\Delta V_{oc}^{nr}$ from growing again after the parabola reaches its minimum. We stress that this optimistic-yet-realistic model (illustrated by the red curve in **Figure 4a**) has no underlying theoretical framework and should not be taken as a lower limit for $\Delta V_{oc}^{nr}$ in OPVs – it is just a means for encapsulating the general trend shown by the experimental data in Figure 4a. For comparison, another semi-analytical model for $\Delta V_{oc}^{nr}$ based on the work of Benduhn et al. is included in Figure 4a. [48] In this model, where a negligibly-small reorganization energy has been assumed, $\Delta V_{oc}^{nr}$ relates to $E_{CT}$ via the so-called energy gap law:

$$\Delta V_{oc}^{nr} \approx C - DE_{CT} \approx C - DE_{opt}, \quad \textbf{(14)}$$

where, $C = 0.574$ V and $D = 0.184 \text{ V eV}^{-1}$. We note that more complex models have been detailed in the literature, including the work of Azzouzi et al. and Chen et al. [41, 52]

The effect of the two non-radiative open-circuit voltage loss models on the open-circuit voltage and PCE are illustrated in **Figure 4b** and **Figure 4c**, respectively. To simulate these curves, a step-function $EQE_{PV}$ was used with $EQE_{max} = 0.85$. It is evident from these curves that accounting for realistic non-radiative open-circuit voltage losses reduces the maximum PCE from 45% to around 40%, while blue-shifting the highest-performing optical gap from 1.83 eV to 1.88 eV. Comparable results are produced by both the semi-analytical energy gap law model given by Equation **(14)** and the optimistic, empirical model given by Equation **(13)**. However, as the PCE differs by just a few percent between the models, we herein utilize Equation **(13)** to model non-radiative losses to make an optimistic prediction for IPV performance.

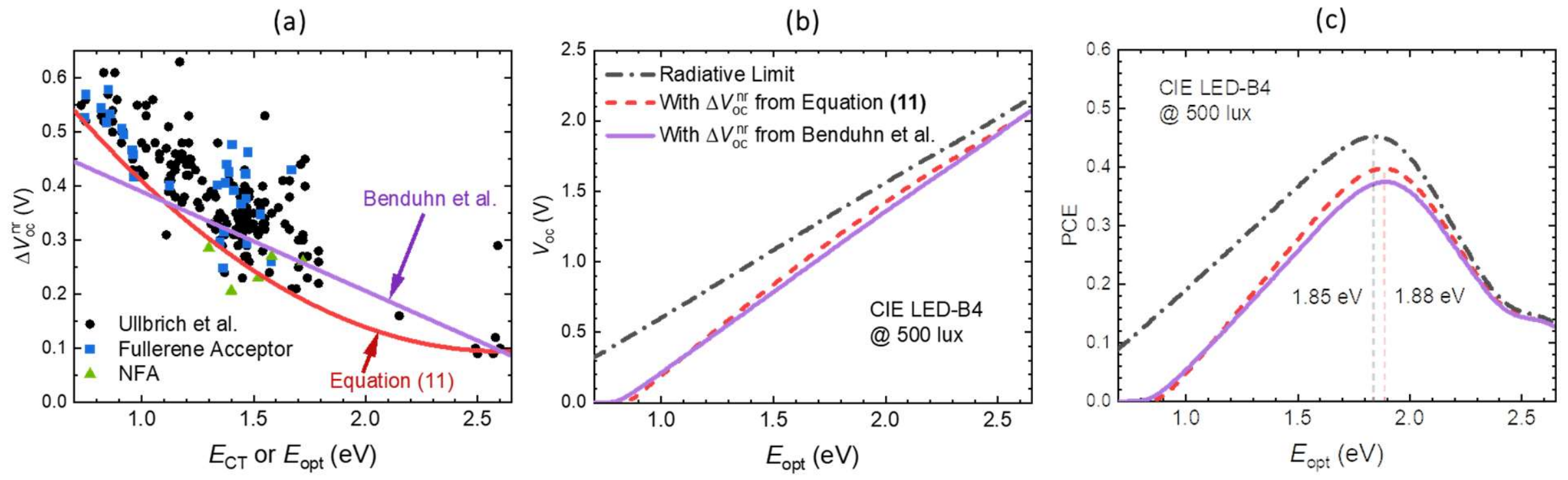

**Figure 4**: The effect of non-radiative open-circuit voltage losses on the PCE of indoor photovoltaics. **(a)** Non-radiative open-circuit losses as a function of the energy of the CT state, $E_{\mathrm{CT}}$, with experimental data compiled by Ullbrich et al. plotted as black squares.[39] Additional data points for OPV blends with fullerene acceptors and NFAs are plotted as blue squares and green triangles, respectively. The empirical model for $\Delta V_{\mathrm{oc}}^{\mathrm{nr}}$ given by Equation **(13)** is indicated by the solid red curve, while Benduhn et al.'s empirical model given in Equation **(14)** is indicated by the purple curve.[48] **(b)** The open-circuit voltage against the optical gap in the radiative limit (black dash-dot curve) and in two non-radiative open-circuit voltage loss models (red dashed curve for Equation **(13)** and solid purple curve for Equation **(14)**) calculated using the step function model for $\mathrm{EQE_{PV}}$ with $\mathrm{EQE_{max}} = 0.85$. **(c)** The resultant power conversion efficiencies under the CIE LED-B4 spectrum at $500$ lux.

The predicted PCEs of organic semiconductor-based IPVs, accounting for both radiative losses and non-radiative losses, are shown in **Figure 5** for the CIE LED-B4 spectrum. The OPV predictions (for both $\sigma_{\mathrm{s}} = 0$ and $\sigma_{\mathrm{s}} = 50$ meV) assume sub-gap absorption calculated using Equation **(11)** and additional $\Delta V_{\mathrm{oc}}^{\mathrm{nr}}$ loss given by Equation **(13)**. Note that $\mathrm{EQE_{max}} = 0.85$ was used to predict an optimistic upper limit for OPVs. For comparison, the ideal radiative PCE limits based on the step-function model (Equation **(7)**) with $\mathrm{EQE_{max}} = 1$ (i.e., the SQ model) and the more realistic $\mathrm{EQE_{max}} = 0.85$ have been included to indicate the performance loss across all optical gaps. These five curves are plotted against the optical gap at an illuminance of 500 lux in **Figure 5a**, whereas, in **Figure 5b,** they are plotted against the illuminance of the incident light for the best-performing optical gap (which has been inset into the graph for each curve).

By accounting for sub-gap absorption, energetic disorder, and realistic non-radiative open-circuit voltage losses, we find that the maximum PCE of OPVs under CIE LED-B4 at 500 lux is reduced from its SQ model value of 53% to around 37%. Furthermore, the highest-performing $E_{\text{opt}}$ is blue-shifted by around 90 meV. Corresponding discussions for the 2700K LED and 4000K LED sources are provided in **Section S6** of the **Supporting Information**. Additionally, similar figures for three standard fluorescent sources (CIE FL-2, CIE FL-7, and CIE FL-11) are illustrated in **Figure S8** of the **Supporting Information**. We note again that the simulated results are mostly independent of the source of artificial light.

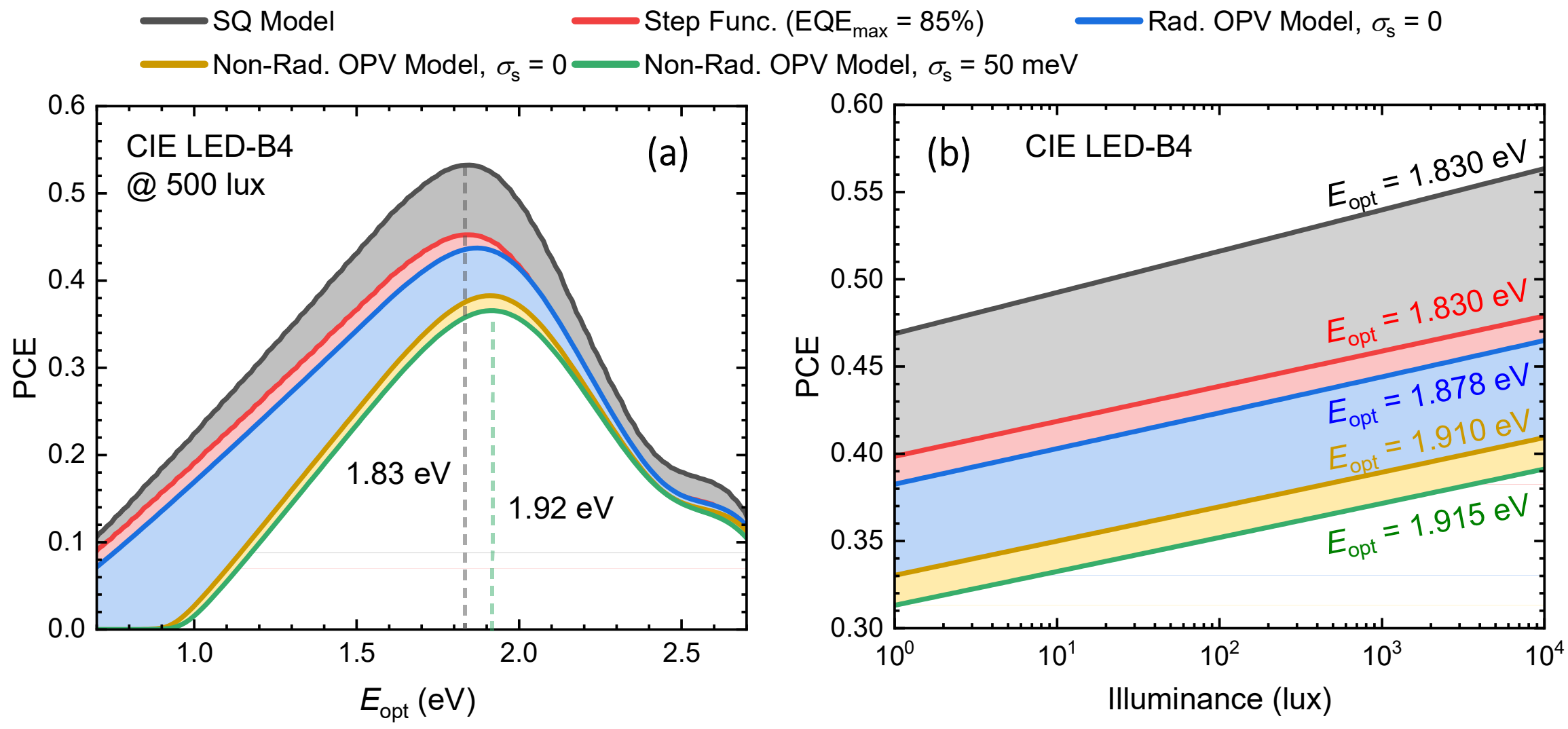

**Figure 5**: Power conversion efficiencies simulated under the CIE LED-B4 spectrum as a function of the optical gap in **(a)**, and as a function of the illuminance in **(b)**. In both panels, five curves are present. The black curves illustrate the PCE in the radiative SQ Model, while the red curves indicate the PCE when $\text{EQE}_{\text{PV}}$ is modelled as a step function with $\text{EQE}_{\text{max}} = 0.85$. The grey shaded regions illustrate the PCE losses induced by non-unity $\text{EQE}_{\text{max}}$. The blue, yellow, and green curves, on the other hand, were simulated with $\text{EQE}_{\text{PV}}$ modelled using Equation **(11)** and $\text{EQE}_{\text{max}} = 0.85$; the blue and yellow curves indicate the $\sigma_s = 0$ case in the radiative and non-radiative limit, respectively, such that the red and blue shaded regions correspond to the losses induced by sub-gap absorption with $E_U = kT$, and non-radiative losses, respectively. In the non-radiative limit, $\Delta V_{oc}^{nr}$ is assigned for a given optical gap using Equation **(13)**. Finally, the green curves indicate the non-radiative limit for $\sigma_s = 50$ meV, with the yellow shaded region indicating the additional loss induced by this disorder. In **(b)**, the highest-performing optical gaps used to simulate the curves are inset.

Based on Figure 5, an OPV blend with the highest-performing optical gap of $E_{opt} = 1.92$ eV and minimal energetic disorder will likely have a PCE lower than 40% at typical indoor light intensities (up to 5000 lux). Accounting for energetic disorder (typically on the order of $\sigma_s = 50$ meV) further reduces the PCE. To estimate the figures-of-merit of particular OPV materials in indoor settings more accurately, we have devised a methodology and created an associated computational tool with an accessible graphical user interface (available as **Supporting Material**) that takes an experimentally-determined $EQE_{PV}$ spectrum and measured open-circuit voltage under one Sun ($V_{oc}^{\odot}$) as inputs. We stress that similar approaches for predicting IPV performance using $EQE_{PV}$ and current-voltage measurements have been established in the past (see, e.g., the work of Lübke et al. [18]). In our case, however, we focus on predicting upper performance limits using measured $EQE_{PV}$ spectra, which account for sub-gap absorption in real OPV systems. Using a device's $EQE_{PV}$ spectrum and $V_{oc}^{\odot}$, the non-radiative open-circuit voltage loss is estimated through

$$\Delta V_{oc}^{nr} \approx \frac{k_B T}{q} \ln\left(1 + \frac{J_{sc}^{\odot}}{J_0^{rad}}\right) - V_{oc}^{\odot}, \quad \textbf{(15)}$$

where $J_{sc}^{\odot}$ is the short-circuit current density under one Sun (determined using $EQE_{PV}$). Assuming Equation **(15)** provides a realistic lower limit estimate of the device's $\Delta V_{oc}^{nr}$, optimistic values for the photovoltaic figures-of-merit can then be estimated under any spectrum at any intensity. A block diagram detailing this methodology, including the identification of the true radiative open-circuit voltage in the thermodynamic limit, is shown in **Figure S9** of the **Supporting Information**. [51]

## 2.4. Comparative Analysis

In **Figure 6a**, the PCE is plotted in the SQ model and in both the radiative limit (blue curve) and non-radiative case (gold curve) of the more realistic OPV model (in the case that $\sigma_s = 0$), where $EQE_{PV}$ is modelled using Equation **(11)**. For the non-radiative case the $\Delta V_{oc}^{nr}$ is assumed to be given by Equation **(13)**. Also shown are the predicted PCEs of state-of-the-art OPVs, [8, 31, 49-51, 53-55] crystalline and amorphous silicon, [51, 56] and a single cation perovskite [54] under CIE LED-B4 (see **Table S5** in the **Supporting Information**). These predictions were made using each system's $EQE_{PV}$ spectrum and $V_{oc}^{\odot}$ from the literature, with the optical gaps taken from the tables of Almora et al.. [49, 50] Where sensitive $EQE_{PV}$ measurements were available, however, the optical gaps were determined using Equation **(11)** via the technique summarized in **Section S8** of the **Supporting Information**. [33]

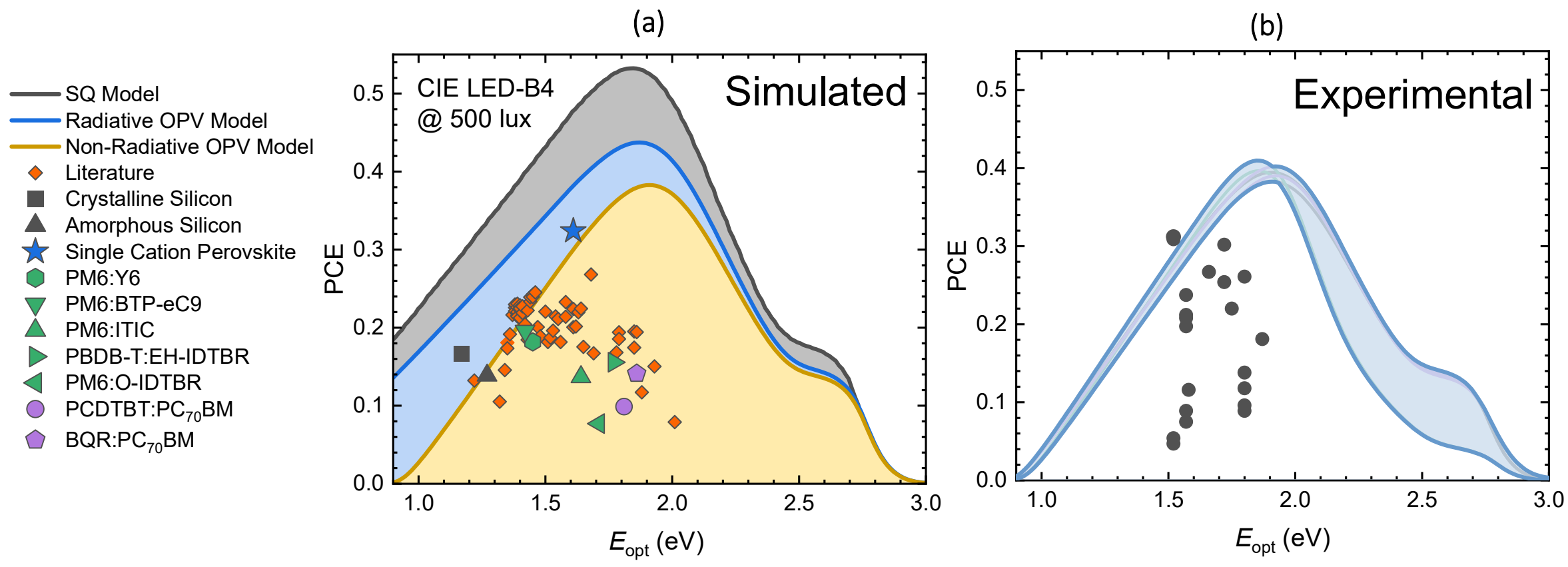

**Figure 6**: Power conversion efficiencies under indoor lighting conditions. (**a**) A comparison of the predicted indoor performance of OPV systems (Almora et al. systems in orange, [49, 50] additional fullerene acceptor and NFA systems in green and purple, respectively), crystalline and amorphous silicon (black data points) and a single cation perovskite (blue star), with the PCE in the SQ model (black curve), and in the radiative and non-radiative OPV predictions in the limit of $\sigma_s = 0$ (shown by the blue and gold curves, respectively). The blue shaded region indicates a regime of optimal performance for materials with low energetic disorder and low $\Delta V_{oc}^{nr}$, whereas the gold shaded region indicates a realistic regime for disordered OPV systems with medium-to-high $\Delta V_{oc}^{nr}$. (**b**) Experimental PCEs under LED sources from the literature (see **Table S6** of the **Supporting Information**), compared with the non-radiative OPV model as a function of the optical gap for the CIE LED-B4, 2700K LED, and 4000K LED sources, at both 500 lux and 2000 lux in the hypothetical $\sigma_\mathrm{s} = 0$ case.

From the estimated PCEs of Figure 6a, it is evident that the state-of-the-art organic solar cell blends PM6:Y6 and PM6:BTP-eC9 (with $E_\mathrm{opt} \sim 1.4$ eV) will likely not exceed PCEs of 20% in indoor settings unless the non-radiative losses can be drastically reduced. As previously discussed in Figure 5, this becomes more evident when accounting for energetic disorder (around 43 meV for both blends) as it further reduces the radiative limit. On the other hand, other OPV systems such as PBDB-T:EH-IDTBR and PM6:O-IDTBR, have a fair amount of room for improvement. This is particularly clear when comparing with experimentally-determined PCE values from the literature, as evidenced in **Figure 6b** (see **Table S5** of the **Supporting Information**; all plotted PCEs were measured at 2000 lux or less). [19, 29, 57-62] Alongside this data, the PCE predicted by the realistic, non-radiative OPV model (with $\sigma_\mathrm{s} = 0$) has been plotted for the CIE LED-B4, 2700K LED, and 4000K LED spectra at both 500 lux and 2000 lux. An envelope has then been plotted to encapsulate the minimum and maximum PCE held by any of the spectra at each optical gap, providing a realistic estimate for PCE of IPVs under *any* LED spectrum.

Based on Figure 6, we can see that many of the OPV blends have room for improvement. We also note, however, that at least one data point from the literature defies the realistic OPV limit with reasonable non-radiative loss, despite the fact that the simulations were conducted at a higher illuminance. Possible reasons for this deviation might be related to i) inaccuracies in the estimated optical gap, ii) a very small non-radiative voltage loss in this OPV system, especially as Equation **(13)** predicts a larger $\Delta V_\mathrm{oc}^\mathrm{nr}$ than some of the literature data in Figure 4a, and/or iii) inaccuracies in the experimental set-up as previously discussed by Lübke et al. [18] Case studies like this demonstrate not only the challenging task of measuring indoor PCEs, but also the dire need for a tried and tested standard for IPVs, including an accepted experimental methodology for characterizing the devices.

## 3. Conclusion

IPVs are rapidly proving to be a very practical application for organic semiconductors; they continue to be a promising contender for powering for the IoT using energy-harvesting techniques. There are, however, some inconsistencies in the literature regarding the PCEs of IPVs, with some devices seemingly surpassing a generous limit; this is likely due to a lack of accepted testing standard for IPVs. By presenting a realistic limit for the PCE of OPVs, which accounts for both radiative open-circuit voltage losses induced by sub-gap absorption (including Urbach tails and energetic disorder) and non-radiative open-circuit voltage losses, we aim to elucidate what PCEs could reasonably be expected. In particular, we have shown that a combination of realistic above-gap $\mathrm{EQE_{PV}}$ and $\Delta V_{\mathrm{oc}}^{\mathrm{nr}}$, in combination with a typical energetic disorder ($\sigma_{\mathrm{s}} = 50$ meV), can reduce the maximum PCE of OPVs from a SQ model value of $53\%$ to $\sim 37\%$ under indoor lighting conditions. We have also shown that the best-performing optical gap becomes blue-shifted from $E_{\mathrm{opt}} = 1.83$ eV to $1.92$ eV in the $\sigma_{\mathrm{s}} = 50$ meV limit of the OPV model, suggesting that the high-performance OPV blends PM6:Y6 and PM6:BTP-eC9 may not exceed PCEs of 20% in indoor settings. To aid future work on indoor applications of OPVs, we have presented a methodology for estimating the performance of IPVs at typical illuminances, using measurements of the photovoltaic external quantum efficiency spectrum and the open-circuit voltage under one Sun. Furthermore, to automate the estimation of IPV performance under arbitrary illumination conditions using these quantities, we have provided a computational tool (with a graphical user interface) as **Supporting Material**.

## 4. The Computational Tool

To aid the estimation of the PCEs of particular photovoltaic materials, a computational tool was prepared in the open-source Python interactive development environment, Jupyter. While this is not the first computational tool for simulating photovoltaic figures-of-merit under indoor illumination conditions, it does have a few unique characteristics. Chief among these, this tool includes a detailed graphical user interface that can be used to control the simulations. To estimate IPV device performance, the tool allows the use of both simulated and experimentally-determined $EQE_{PV}$ spectra; it can simulate step-functions, sub-gap Urbach tails, and OPV absorption using Equation **(11)**. Using these simulated $EQE_{PV}$ spectra, the tool determines the photovoltaic figures-of-merit under a selected spectrum (e.g., CIE LED-B4) at any desired illuminance. A variety of non-radiative open-circuit voltage loss models are also available, including Equation **(13)**. The photon flux spectra used by the tool can be customized (and superimposed), and as many $EQE_{PV}$ spectra as desired can be loaded in. These spectra may be analyzed individually, or countless systems may be analyzed at once in bulk – enabling a prediction of which device would perform best out of a selection of hundreds under a given spectrum in a matter of seconds. The tool is applicable to any class of semiconductor materials, including organics, inorganics, and perovskites. Alongside the well-detailed tool, a manual has also been prepared that describes how to install Jupyter and how to navigate the user interface.

## Acknowledgements

We kindly acknowledge Dr. Nasim Zarrabi, Dr. Stefan Zeiske, Dr. Christina Kaiser, and Dr. Wei Li, for providing experimental photovoltaic external quantum efficiency data. This work was funded through the Welsh Government's Sêr Cymru II Program 'Sustainable Advanced Materials' (Welsh European Funding Office − European Regional Development Fund). P.M. is a Sêr Cymru II Research Chair and A.A. is a Rising Star Fellow also funded through the Welsh Government's Sêr Cymru II 'Sustainable Advanced Materials' Program (European Regional Development Fund, Welsh European Funding Office and Swansea University Strategic Initiative). This work was also funded by the UKRI through the EPSRC Program Grant EP/T028513/1 Application Targeted and Integrated Photovoltaics.

## Data Availability Statement

The data that support the results of this work will be made available upon request from the corresponding authors.

## Conflicts of Interest

The authors declare no conflicts of interest.

Supporting Information

# The Thermodynamic Limit of Indoor Photovoltaics Based on Energetically-Disordered Molecular Semiconductors

*Austin M. Kay[1], Maura E. Fitzsimons[1], Gregory Burwell[1], Paul Meredith[1], Ardalan Armin[1], and Oskar, J. Sandberg[1]*

[1]Sustainable Advanced Materials (Sêr-SAM), Centre for Integrative Semiconductor Materials (CISM), Department of Physics, Swansea University Bay Campus, Swansea SA1 8EN, United Kingdom

Email: ardalan.armin@swansea.ac.uk; o.j.sandberg@swansea.ac.uk

## Table of Contents

# Part I – Supporting Theory and Simulation Results

## S1. Units of Illuminance

To provide an estimate of the upper limit of indoor photovoltaic (IPV) device performance, appropriate units should be used to describe low light intensities; these units are lux (lx) and they quantify the *illuminance* of a source. For a spectral photon flux density $\Phi_{\text{source}}$, the illuminance ($L_{\text{source}}$) is defined as [1-3]

$$L_{\text{source}} = L_0 P_{\text{source}} \int_0^{\infty} V(E) \cdot E \cdot \tilde{\Phi}_{\text{source}}(E)\, \mathrm{d}E = P_{\text{source}} f_{\text{source}}, \quad \textbf{(S1)}$$

where $f_{\text{source}} = L_0 \int_0^{\infty} V(E) \cdot E \cdot \tilde{\Phi}_{\text{source}}(E)\, \mathrm{d}E$ contains all the spectral information, $L_0 = 683\ \text{lx} \cdot \text{W}^{-1} \cdot \text{m}^2$ is a constant, $P_{\text{source}} = \int_0^{\infty} E \cdot \Phi_{\text{source}}(E)\, \mathrm{d}E$ is the integrated irradiance of the source, and $\tilde{\Phi}_{\text{source}}$ is the spectral photon flux normalized to $P_{\text{source}}$. The luminous efficiency for photopic vision, $V(E)$, is plotted as a function of the photon energy ($E$) at a $2^{\circ}$ viewing angle in **Figure S1a**. [2]

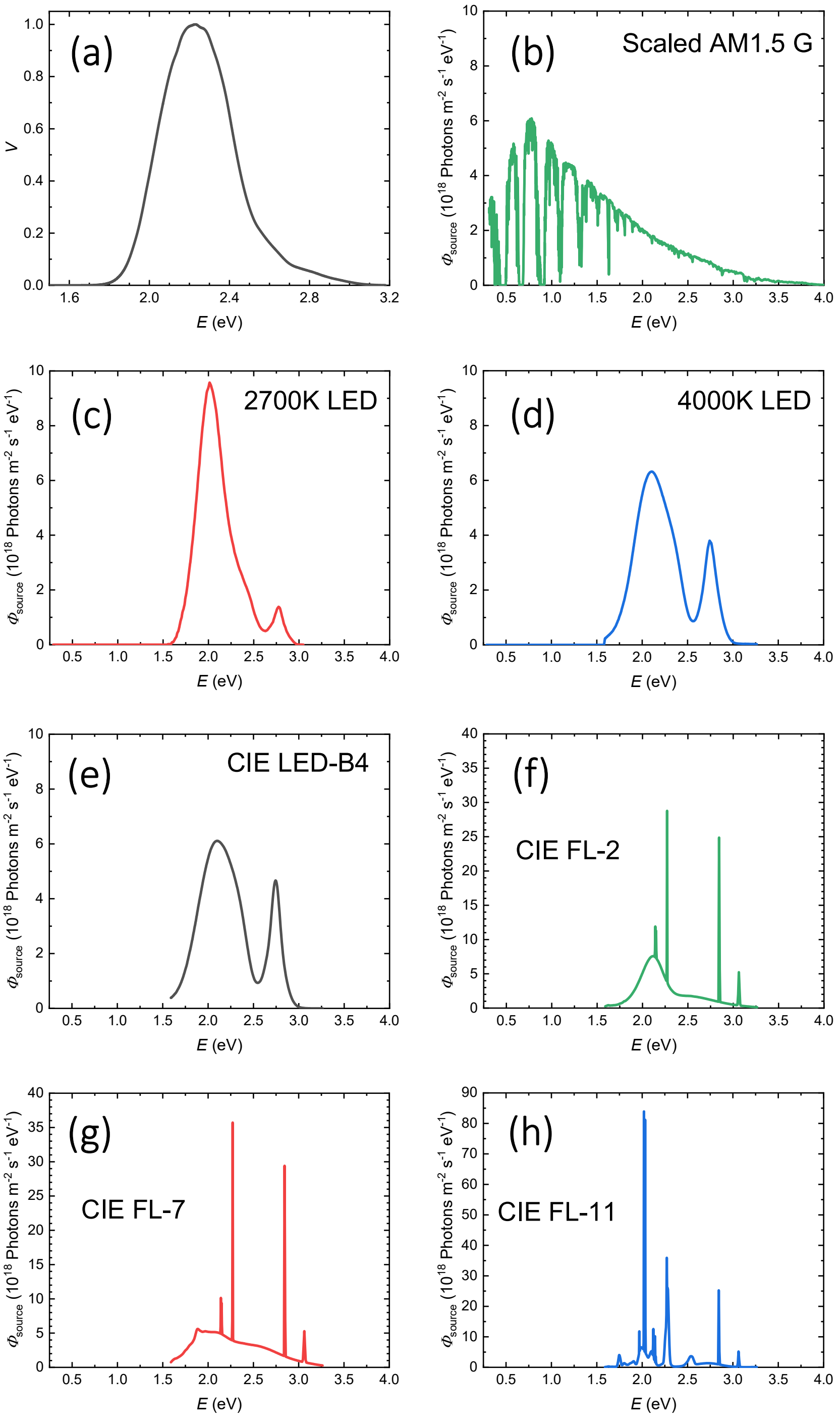

S

**Figure S1**: Converting to units of illuminance. (**a**) The luminous efficiency $V(E)$ for photopic vision at a 2° viewing angle, plotted as a function of the photon energy, $E$. [2] (**b-h**) The spectral photon flux densities

for a variety of light sources at an illuminance of 500 lux, including the scaled AM1.5G for sunlight, 2700K LED, 4000K LED, and the standard illuminant CIE LED-B4, CIE FL-2, CIE FL-7, and CIE FL-11 spectra. Note that the scaled AM1.5G and LED spectra share the same vertical axis, while the others have larger scales.

Using the spectral luminous efficiency of **Figure S1a**, the relationship between illuminance and irradiance was determined for a variety of spectra, including the 2700K LED, 4000K LED, AM1.5G, and the standard illuminants CIE LED-B4, CIE FL-2, CIE FL-7, and CIE FL-11; [4, 5] the results are compiled in **Table S1**. Using these values, the photon fluxes associated with each of these spectra at an illuminance of 500 lux has been plotted in **Figure S1b**.

**Table S1**: Relationship between lux and irradiance for different spectra, determined using Equation **(S1)** and the luminous efficiency data plotted in **Figure S1a**.

| **Spectrum** | $f_{\text{source}}$ $(\text{lx} \cdot \text{W}^{-1} \cdot \text{m}^2)$ |
|---|---|
| AM1.5G | 115.61 |
| 2700K LED | 349.30 |
| 4000K LED | 346.13 |
| CIE LED-B4 | 333.62 |
| CIE FL-2 | 355.26 |
| CIE FL-7 | 268.19 |
| CIE FL-11 | 356.68 |

## S2. Figures of Merit in the Ideal Diode Model

Neglecting series and shunt resistances and assuming a unity ideality factor, the total current density generated under illumination ($J_{\mathrm{light}}$) by an ideal Shockley diode (with unity ideality factor) is given by [6, 7]

$$J_{\mathrm{light}} = J_0\left[\exp\left(\frac{qV}{kT}\right) - 1\right] - J_{\mathrm{sc}}, \tag{S2}$$

where $V$ is the voltage applied to the diode, $k$ is the Boltzmann constant, $T$ is the temperature, $J_{\mathrm{sc}} = q\int_0^\infty \mathrm{EQE}_{\mathrm{PV}}(E)\Phi_{\mathrm{source}}(E)\,\mathrm{d}E$ is the short-circuit current density under illumination, and $J_0 = \frac{q}{\mathrm{EQE}_{\mathrm{EL}}}\int_0^\infty \mathrm{EQE}_{\mathrm{PV}}(E)\Phi_{\mathrm{bb}}(E)\,\mathrm{d}E$ is the dark saturation current density. In these expressions, $q$ denotes the elementary charge, $\Phi_{\mathrm{bb}}$ is the spectral photon flux associated with a black body at temperature $T$, and $\mathrm{EQE}_{\mathrm{PV}}$ and $\mathrm{EQE}_{\mathrm{EL}}$ are the photovoltaic and electroluminescent external quantum efficiencies, respectively.

At open-circuit conditions, where the applied voltage is equal to the open-circuit voltage ($V_{\mathrm{oc}}$), no net current is produced by the device. Rearranging Equation **(S2)** in this case yields

$$V_{\mathrm{oc}} = \frac{kT}{q}\ln\left[1 + \frac{J_{\mathrm{sc}}}{J_0}\right]. \tag{S3}$$

To calculate the power conversion efficiency (PCE), the power outputted by the device at the maximum power point ($V_{\mathrm{mpp}}$) must be determined. This is done by first multiplying Equation **(S2)** with $V$ to find the power density. Following this, taking the derivative with respect to the applied voltage, setting the whole expression equal to nought, and rearranging gives $\exp\left(1 + \frac{qV_{\mathrm{oc}}}{kT}\right) = \left(1 + \frac{qV_{\mathrm{mpp}}}{kT}\right)\exp\left(1 + \frac{qV_{\mathrm{mpp}}}{kT}\right)$, which was solved in this work using the Lambert $W$ function, as seen in other works in the literature. [8-11] The maximum power point voltage then relates to $V_{\mathrm{oc}}$ via

$$V_{\mathrm{mpp}} = \frac{kT}{q}\left(W\left[\exp\left(1 + \frac{qV_{\mathrm{oc}}}{kT}\right)\right] - 1\right). \tag{S4}$$

The solar cell's fill factor (FF) was then determined using:

$$\mathrm{FF} = \frac{|J_{\mathrm{mpp}}|V_{\mathrm{mpp}}}{J_{\mathrm{sc}}V_{\mathrm{oc}}}, \tag{S5}$$

where the current density at the maximum power point ($J_{\mathrm{mpp}}$) is determined by substituting $V_{\mathrm{mpp}}$, evaluated using Equation **(S4)**, back into Equation **(S2)**.

A free-standing expression for the fill factor in terms of the open-circuit voltage can be obtained from Equations **(S1)** to **(S5)** using Newton's method to approximate the Lambert $W$ function (in the large-argument limit) as [12]

$$W[x] \approx \ln(x)\left(1 - \frac{\ln[\ln(x)]}{1+\ln[x]}\right). \qquad \textbf{(S6)}$$

Consequently, the maximum power point voltage can be written in terms of the open-circuit voltage as

$$V_{\mathrm{mpp}} \approx V_{\mathrm{oc}}\left[1 - \frac{kT}{qV_{\mathrm{oc}}}\left(\frac{1+\frac{qV_{\mathrm{oc}}}{kT}}{2+\frac{qV_{\mathrm{oc}}}{kT}}\right)\ln\left[1+\frac{qV_{\mathrm{oc}}}{kT}\right]\right]. \qquad \textbf{(S7)}$$

Whereas the photocurrent density at the maximum power point in terms of the open-circuit voltage is, with $\frac{J_0}{J_{\mathrm{sc}}} = \left[\exp\left(\frac{qV_{\mathrm{oc}}}{kT}\right) - 1\right]^{-1}$,

$$J_{\mathrm{mpp}} = J_{\mathrm{sc}}\left[\exp\left(-\left[\frac{1+\frac{qV_{\mathrm{oc}}}{kT}}{2+\frac{qV_{\mathrm{oc}}}{kT}}\right]\ln\left[1+\frac{qV_{\mathrm{oc}}}{kT}\right]\right) - 1\right]\left[1-\exp\left(-\frac{qV_{\mathrm{oc}}}{kT}\right)\right]^{-1}. \qquad \textbf{(S8)}$$

The fill factor can then be written as

$$\mathrm{FF} \approx \frac{\left[1 - \frac{kT}{qV_{\mathrm{oc}}}\left(\frac{1+\frac{qV_{\mathrm{oc}}}{kT}}{2+\frac{qV_{\mathrm{oc}}}{kT}}\right)\ln\left[1+\frac{qV_{\mathrm{oc}}}{kT}\right]\right]\left[1-\exp\left(-\left[\frac{1+\frac{qV_{\mathrm{oc}}}{kT}}{2+\frac{qV_{\mathrm{oc}}}{kT}}\right]\ln\left[1+\frac{qV_{\mathrm{oc}}}{kT}\right]\right)\right]}{1-\exp\left(-\frac{qV_{\mathrm{oc}}}{kT}\right)}. \qquad \textbf{(S9)}$$

For open-circuit voltages larger than around 0.2 V, this reduces to

$$\mathrm{FF} \approx \left[1 - \frac{kT}{qV_{\mathrm{oc}}}\left(\frac{1+\frac{qV_{\mathrm{oc}}}{kT}}{2+\frac{qV_{\mathrm{oc}}}{kT}}\right)\ln\left[1+\frac{qV_{\mathrm{oc}}}{kT}\right]\right]\left[1-\exp\left(-\left[\frac{1+\frac{qV_{\mathrm{oc}}}{kT}}{2+\frac{qV_{\mathrm{oc}}}{kT}}\right]\ln\left[1+\frac{qV_{\mathrm{oc}}}{kT}\right]\right)\right]. \qquad \textbf{(S10)}$$

In the further limit that the open-circuit voltage is greater than around 0.5 V, Equation **(S10)** can be further simplified to give

$$\mathrm{FF} \approx \frac{\frac{qV_{\mathrm{oc}}}{kT} - \ln\left(1 + \frac{qV_{\mathrm{oc}}}{kT}\right)}{1 + \frac{qV_{\mathrm{oc}}}{kT}}. \quad \textbf{(S11)}$$

This expression for the fill factor is the same as in Würfel's *Physics of Solar Cells*.[13] In the remainder of this work (and in the computational tool we provide as **Supporting Material**), we calculate the fill factor through the maximum power point voltage given by Equation **(S4)** and the current density at the maximum power point, which is determined by substituting $V = V_{\mathrm{mpp}}$ into Equation **(S2)**. By doing this, we can more readily compute the maximum power point parameters, which is ideal as they are more pertinent for indoor applications than figures-of-merit like the fill factor and the power conversion efficiency. We stress that Equation **(S11)** is valid for the open-circuit voltages of most working solar cells, and most indoor photovoltaics. As shown by the deviation ($\Delta\mathrm{FF} = |\mathrm{FF}_{\mathrm{num}} - \mathrm{FF}_{\mathrm{an}}|/\mathrm{FF}_{\mathrm{num}}$) between the numerically-calculated fill factor ($\mathrm{FF}_{\mathrm{num}}$) in the Shockley-Queisser (SQ) model and the fill factor calculated analytically ($\mathrm{FF}_{\mathrm{an}}$) using Equation **(S9)** to **(S11)** in **Figure S2**, Equation **(11)** is a good approximation for $V_{\mathrm{oc}} > 0.5$ V.

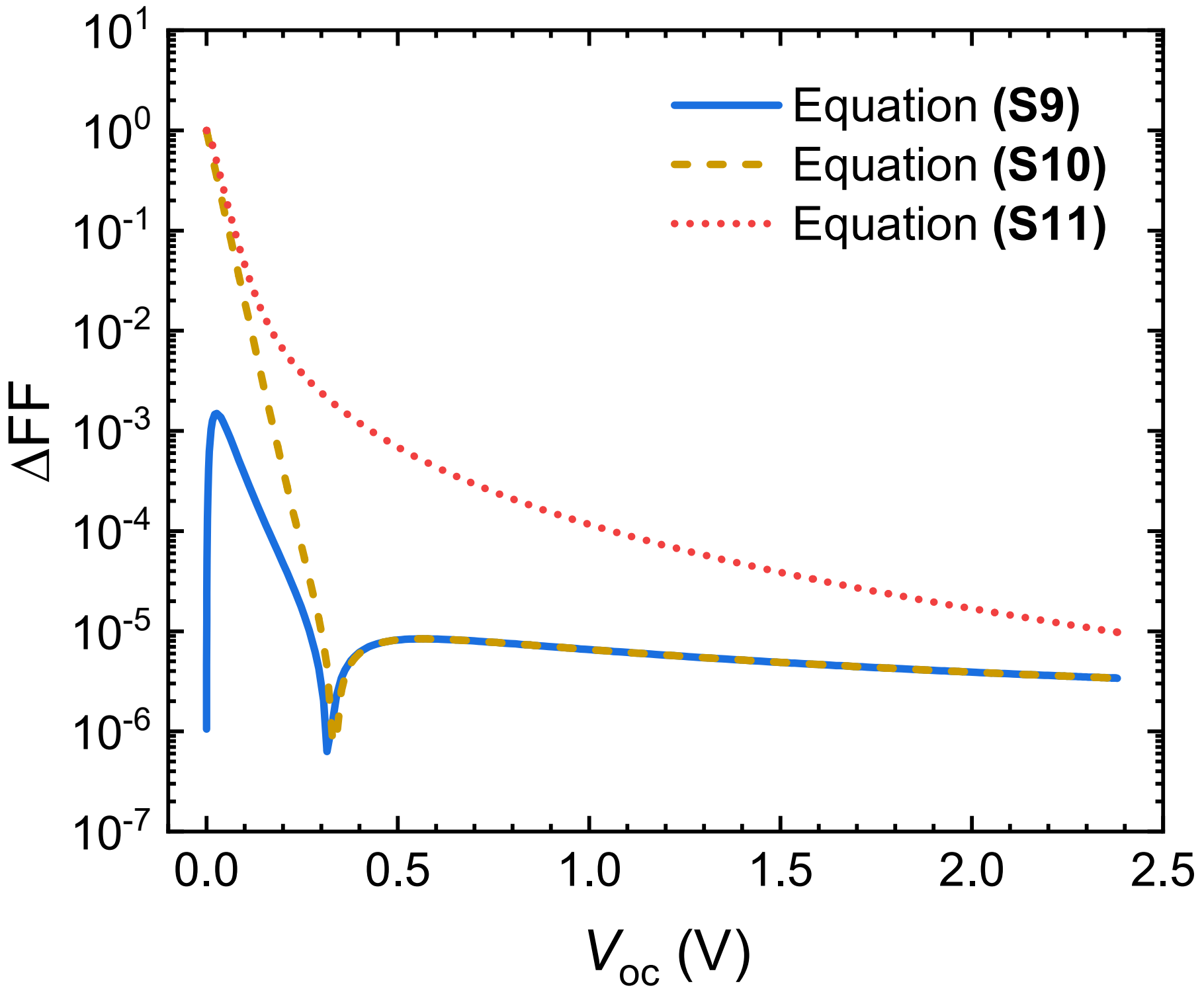

**Figure S2:** Deviation between the numerically-calculated and analytically-approximated fill factor (ΔFF) in the Shockley-Queisser model. The deviation in the case that the FF is modelled using Equation **(S9)** is shown by the solid, blue curve, while the deviations when the FF is calculated using Equation **(S10)** and Equation **(S11)** are illustrated by the dashed, golden curve and dotted, red curve, respectively. These curves show that Equation **(S11)** is a good approximation for $V_{\mathrm{oc}} > 0.5$ V.

## S3. Incorporating State-Filling Effects

To accurately model photovoltaic performance in devices with sub-gap absorption, state-filling effects need to be accounted for as recently pointed out by Wong et al. [14] State filling generally becomes important for states with energies comparable or below the chemical potential ($\mu$) of the emissive species, necessitating the use of the generalized Planck radiation for emission; [13, 15] in the thermodynamic limit, $\mu$ is equal to the quasi-Fermi level splitting of free charge carriers with the radiative current density given by

$$J_{\mathrm{rad}}(V) = q\int_0^{\infty} \mathrm{EQE}_{\mathrm{PV}}(E,V)\Phi_{\mathrm{em}}(E,V)\,\mathrm{d}E, \quad \mathbf{(S12)}$$

where

$$\Phi_{\mathrm{em}}(E,V) = \frac{2\pi E^2}{h^3c^2}\frac{1}{\exp\left(\frac{E-qV}{kT}\right)-1}. \quad \mathbf{(S13)}$$

Here, $h$ is the Planck constant and $c$ is the speed of light. Assuming the quasi-Fermi level splitting of free charge carriers to be equal to the applied voltage $V$. Furthermore, state filling reduces the number of available excitable states giving rise to a modification of the absorption coefficient $\alpha$ (see Chapter 3.7 of Refs [13, 15]). Taking $\mathrm{EQE}_{\mathrm{PV}} \propto \alpha$ for sub-gap (weak) absorption, we can then approximate

$$\mathrm{EQE}_{\mathrm{PV}}(E,V) = \mathrm{EQE}_{\mathrm{PV}}(E) \times \Delta f(E,V) \quad \mathbf{(S14)}$$

for $E \gg kT$, where $\mathrm{EQE}_{\mathrm{PV}}(E)$ corresponds to the photovoltaic external quantum efficiency at short-circuit ($V = 0$), while $\Delta f(E,V) = f_v - f_c$ is the difference between the (Fermi-Dirac) electron occupation probability in the valence and conduction level. Assuming flat-band conditions with symmetric quasi-Fermi level splitting (relative to the middle of the gap) to prevail, $\Delta f$ is given by [14]

$$\Delta f(E,V) = \tanh\left(\frac{E-qV}{4kT}\right). \quad \mathbf{(S15)}$$

Combining everything together yields

$$J_{\mathrm{rad}}(V) = J_0^{\mathrm{rad}}(V)\exp\left(\frac{qV}{kT}\right). \tag{S16}$$

where we have defined the (effective) dark saturation current density as

$$J_0^{\mathrm{rad}}(V) \approx q\int_0^{\infty} \frac{\mathrm{EQE_{PV}}(E)\Phi_{\mathrm{bb}}(E)\tanh\left(\frac{E-qV}{4kT}\right)}{1-\exp\left(\frac{qV-E}{kT}\right)}\mathrm{d}E = q\int_0^{\infty}\frac{\mathrm{EQE_{PV}}(E)\Phi_{\mathrm{bb}}(E)}{\left[1+\exp\left(\frac{qV-E}{2kT}\right)\right]^2}\mathrm{d}E, \tag{S17}$$

with $\Phi_{\mathrm{bb}}(E)$ corresponding to the black-body radiation at thermal equilibrium ($V = 0$). Note that the assumption that sub-gap tail states with energy $E < qV$ do not contribute to the short-circuit current density may be made, such that state-filling effects may be neglected in the calculation of $J_{\mathrm{sc}}$.

In this work, a binary search algorithm was used to numerically determine the maximum power point parameters and the photovoltaic figures-of-merit for a given $\mathrm{EQE_{PV}}$ spectrum and set of illumination conditions in the case that band-filling effects are present. This iterative approach starts with an evaluation of the radiative, non-radiative, and maximum power point voltage in the case that no state-filling effects are present (as outlined in **Section S2**). Twice these voltages are taken as the initial value for the binary search algorithm, with an upper limit of four times these voltages and a lower limit of nought assumed in each case. Hops towards the solutions are then made by jumping halfway towards the upper or lower limit (depending on whether the total current is negative or positive in the evaluation of the open-circuit voltage, and whether or not the power is increasing or decreasing about the estimated maximum power point voltage). Convergence is reached in, e.g., the evaluation of the open-circuit voltage when the following criterion is met

$$\frac{\left|V_{\mathrm{oc},i} - V_{\mathrm{oc},i+1}\right|}{V_{\mathrm{oc},i}} < \xi, \tag{S18}$$

where $i$ denotes the iteration number, and we take $\xi = 10^{-6}$ for convergence to be reached.

## S4. Models for Sub-Gap Absorption

In the main text, three different models for $\mathrm{EQE_{PV}}$ are considered. The first of these is a rudimentary step function, where photons of greater than the optical gap ($E_{\mathrm{opt}}$) generate an electron-hole pair with efficiency $\mathrm{EQE_{max}}$, while photons of energy less than the optical gap do not:

$$\mathrm{EQE_{PV}}(E) = \begin{cases} \mathrm{EQE_{max}}, & \text{if} \quad E \geq E_{\mathrm{opt}}, \\ 0, & \text{otherwise.} \end{cases} \quad \textbf{(S19)}$$

In the second model, previously proposed to describe excitonic sub-gap absorption for organic photovoltaics (OPVs) by the authors, the $\mathrm{EQE_{PV}}$ is given by[16, 17]

$$\mathrm{EQE_{PV}}(E) = \frac{\mathrm{EQE_{max}}}{2}\left\{\exp\left[\frac{E - E_{\mathrm{opt}} + \frac{\sigma_s^2}{2kT}}{kT}\right]\mathrm{erfc}\left[\frac{E - E_{\mathrm{opt}} + \frac{\sigma_s^2}{kT}}{\sigma_s\sqrt{2}}\right] + \mathrm{erfc}\left[\frac{E_{\mathrm{opt}} - E}{\sigma_s\sqrt{2}}\right]\right\}. \quad \textbf{(S20)}$$

Here, $E_{\mathrm{opt}}$ is the mean optical gap of a Gaussian distribution of exciton states with standard deviation $\sigma_s$ – the excitonic static disorder. In this expression, erfc denotes the complementary error function.[18] The third model for $\mathrm{EQE_{PV}}$ which was considered can be thought of as being halfway between the step function model of Equation **(S19)** and the disorder-dependent $\mathrm{EQE_{PV}}$ model given by Equation **(S20)**. It combines an above-gap quantum efficiency $\mathrm{EQE_{max}}$ with a sub-gap Urbach tail characterized with Urbach energy $E_{\mathrm{U}}$, give[19]

$$\mathrm{EQE_{PV}}(E) = \mathrm{EQE_{max}}\begin{cases} 1, & \text{if} \quad E \geq E_{\mathrm{opt}}, \\ \exp\left(\frac{E - E_{\mathrm{opt}}}{E_{\mathrm{U}}}\right), & \text{otherwise.} \end{cases} \quad \textbf{(S21)}$$

The spectral behavior of Equation **(S20)** and Equation **(S21)** are plotted in **Figure S3a** and **S3b**, respectively.

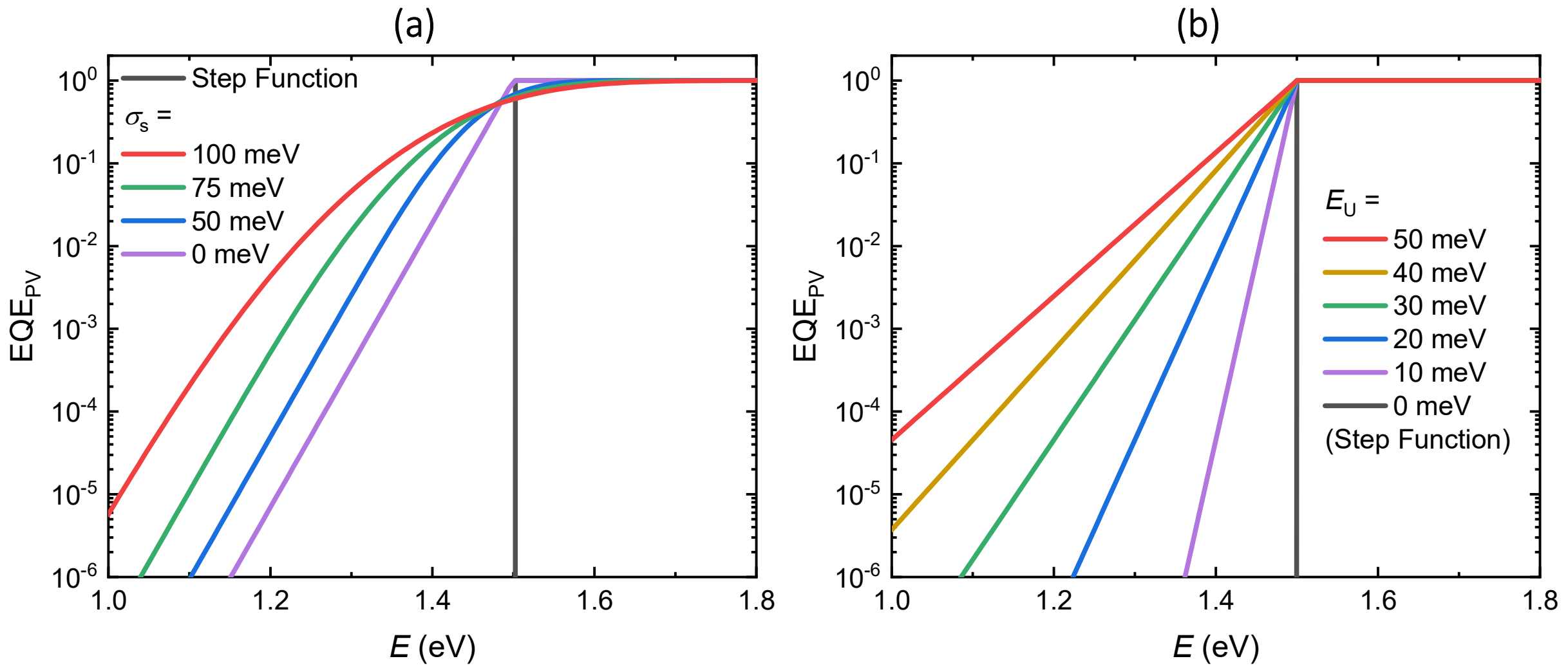

**Figure S3**: Models for the sub-gap photovoltaic external quantum efficiency for $E_{\mathrm{opt}} = 1.5$ eV. (**a**) $\mathrm{EQE_{PV}}$ determined using Equation **(S20)** for a variety of $\sigma_{\mathrm{s}}$ and compared with the step function model determined using Equation **(S19)**. **(b)** $\mathrm{EQE_{PV}}$ in the sub-gap Urbach tail model, for a variety of $E_{\mathrm{U}}$, determined using Equation **(S21)**. The $E_{\mathrm{U}} = 0$ case is equivalent to the step function model.

From **Figure S3**, it is clear that sub-gap Urbach tails with larger $E_{\mathrm{U}}$ have increased sub-gap absorption. The band edges of systems with smaller Urbach energies, like perovskites, are therefore much better defined than the band edges of non-crystalline semiconductors. [20] Using these models for $\mathrm{EQE_{PV}}$ spectra, the dark saturation current densities in the radiative limit (and the consequent open-circuit voltages and power conversion efficiencies) are calculated in the next section.

## S5. Dark Saturation Current Densities

The dark saturation current density in the radiative limit ($J_0^{\mathrm{rad}}$) is determined using the black-body photon flux density, which itself is given by [6, 21, 22]

$$\Phi_{\mathrm{bb}}(E) \approx \frac{2\pi E^2}{h^3 c^2} \exp\left(-\frac{E}{kT}\right). \tag{S22}$$

for $E > 2kT$. [13] Using Equation **(S22)**, in conjunction with the results from **Section S3**, the dark saturation current in the radiative limit is given by

$$J_0^{\mathrm{rad}}(V_{\mathrm{oc}}) \approx \frac{2\pi q}{h^3 c^2} \int_{E_*}^{\infty} \mathrm{EQE_{PV}}(E) E^2 \exp\left(-\frac{E}{kT}\right) \mathrm{d}E, \tag{S23}$$

where the open-circuit voltage dependence has been approximated with a unit step-function at $E = E_*$ assuming $E_* \approx qV_{\mathrm{oc}}$.

The radiative dark saturation current density in the step-function $\mathrm{EQE_{PV}}$ model (assuming $E_{\mathrm{opt}} > E_*$) will be given by substituting Equation **(S19)** into Equation **(S23)** and using the definition of the gamma function, $\Gamma(n+1) = \int_0^{\infty} x^n e^{-x}\,\mathrm{d}x = n!$, giving [18]

$$J_{0,\mathrm{step}}^{\mathrm{rad}} \approx \frac{2\pi q \mathrm{EQE_{max}}}{h^3 c^2} \left[2k^3T^3 + 2E_{\mathrm{opt}}k^2T^2 + E_{\mathrm{opt}}^2 kT\right] \exp\left(-\frac{E_{\mathrm{opt}}}{kT}\right). \tag{S24}$$

The fact that $E_{\mathrm{opt}} \gg kT$ may be used to simplify Equation **(S24)**, but, for now, this approximation is not made. On the other hand, it is shown in the **Appendix** that, by substituting Equation **(S20)** into Equation **(S23)**, the radiative dark saturation current density is given by

$$J_0^{\mathrm{rad}} = f_w \frac{\pi q \mathrm{EQE_{max}}}{h^3 c^2} \left[f_1 \exp\left(-\frac{\Delta^2}{2\sigma_s^2}\right) + f_2 \,\mathrm{erfc}\left(\frac{\Delta}{\sigma_s\sqrt{2}}\right)\right] \exp\left(\frac{-E_{\mathrm{opt}} + \frac{\sigma_s^2}{2kT}}{kT}\right). \tag{S25}$$

Where the parameters $\Delta$, $f_1$, and $f_2$ are defined as

$$\Delta = -E_{\mathrm{opt}} + \frac{\sigma_s^2}{kT} \tag{S26a}$$

$$f_1 = \sigma_s \sqrt{\frac{2}{\pi}} \left( \frac{\Delta^2 + 2\sigma_s^2}{3} - \Delta kT + 2k^2T^2 \right), \tag{S26b}$$

$$f_2 = (\Delta^2 + \sigma_s^2)kT - \Delta(2k^2T^2 + \sigma_s^2) + 2k^3T^3 - \frac{\Delta^3}{3}, \tag{S26c}$$

while $f_\mathrm{w}$ is a correction factor due to state-filling. In the limit that $E_\mathrm{opt} \gg \sigma_\mathrm{s}$, the exponential term decays rapidly and the complementary error function term tends to a constant value of two, giving

$$\begin{aligned} J_0^\mathrm{rad}\Big|_{E_\mathrm{opt} \gg \sigma_\mathrm{s}} &\approx f_w \frac{2\pi q \mathrm{EQE}_\mathrm{max} f_2}{h^3c^2} \exp\left( \frac{-E_\mathrm{opt} + \frac{\sigma_\mathrm{s}^2}{2kT}}{kT} \right) \\ &\approx f_w \left( E_\mathrm{opt} - \frac{\sigma_\mathrm{s}^2}{kT} \right)^3 \frac{2\pi q \mathrm{EQE}_\mathrm{max}}{3h^3c^2} \exp\left( \frac{-E_\mathrm{opt} + \frac{\sigma_\mathrm{s}^2}{2kT}}{kT} \right). \end{aligned} \tag{S27}$$

In the sub-gap Urbach tail model, $\mathrm{EQE_{PV}}$ is instead modelled using Equation **(S21)**, giving the following radiative dark saturation current density

$$J_0^\mathrm{rad} \approx \frac{2\pi q \mathrm{EQE}_\mathrm{max}}{h^3c^2} \left[ h(E_*) - h(E_\mathrm{opt}) \right] \exp\left( -\frac{E_\mathrm{opt}}{E_\mathrm{U}} \right) + J_\mathrm{0,step}^\mathrm{rad}, \tag{S28}$$

with

$$h(E) = \left( \frac{E^2}{m} + \frac{2E}{m^2} + \frac{2}{m^3} \right) e^{-mE} \tag{S29}$$

for $m = \frac{1}{kT} - \frac{1}{E_\mathrm{U}} \neq 0$. Equation **(S28)** is valid only for $E_\mathrm{U} \neq kT$, i.e., for $m \neq 0$. In the special case that $E_\mathrm{U} = kT$ $(m = 0)$, one finds

$$J_0^\mathrm{rad,} \approx \frac{2\pi q \mathrm{EQE}_\mathrm{max}}{h^3c^2} \left[ \frac{E_\mathrm{opt}^3}{3} - \frac{E_*^3}{3} \right] \exp\left( -\frac{E_\mathrm{opt}}{kT} \right) + J_\mathrm{0,step}^\mathrm{rad}. \tag{S30}$$

Both sub-gap models for $\mathrm{EQE_{PV}}$ include additional contributions to the dark saturation current density, which are not present in the simple step function model. As illustrated by the simulated dark saturation current densities of **Figure S4**, these contributions result in orders of magnitude increases. Furthermore, increasing the excitonic disorder $\sigma_s$ is shown to also increase the dark saturation current density, but have little effect on the gradient of the line as a function of $E_{opt}$. On the other hand, increasing $E_U$ beyond 20 meV results in vast increases in $J_0^{rad}$, with the gradients of the curves depending on $E_U$. In panels **a** and **b** of **Figure S4**, it is shown that the short-circuit current density is mostly unperturbed by changes in $\sigma_s$ and $E_U$, whereas variations in the dark saturation current density are by orders of magnitude.

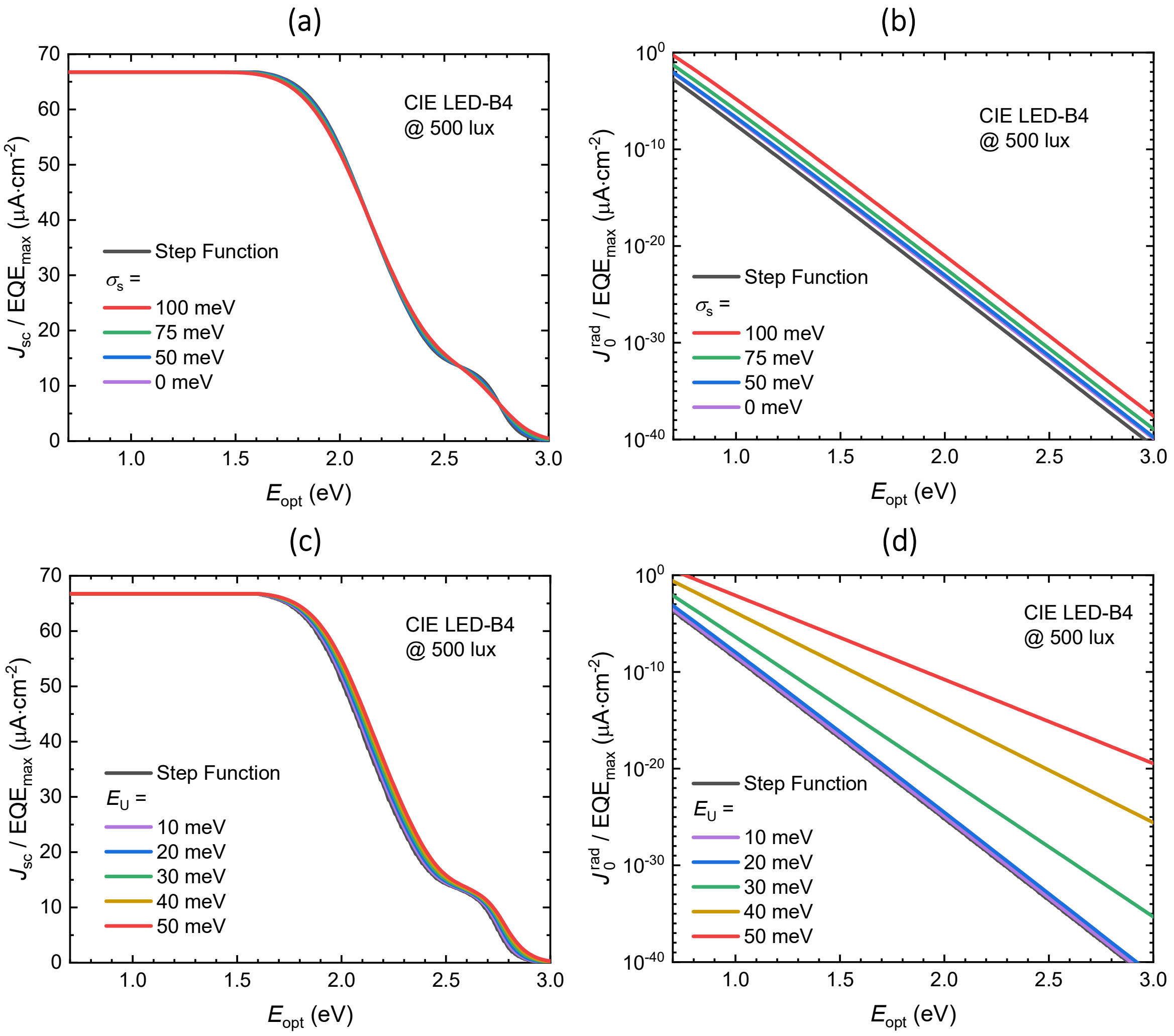

**Figure S4**: Simulated current densities under the CIE LED-B4 spectrum at 500 lux, normalized to the above-gap photovoltaic quantum efficiency, $EQE_{max}$. **(a)** The short-circuit current density and **(b)** the radiative dark saturation current density, both as a function of the optical gap for $\sigma_s$ varied from 0 to 100 meV. These curves were simulated using an $\mathrm{EQE_{PV}}$ spectrum modelled by Equation **(S20)**. Alternatively, in **(c)** and **(d)**, the short-circuit current density and radiative dark saturation current density was simulated using an $\mathrm{EQE_{PV}}$ modelled by Equation **(S21)**, where $E_\mathrm{U}$ has been varied from 0 to 50 meV.

Finally, the radiative open-circuit voltage loss induced by sub-gap absorption is taken as the deviation between the SQ open-circuit voltage ($V_\mathrm{oc}^\mathrm{SQ}$) and the obtained radiative open-circuit voltage ($V_\mathrm{oc}^\mathrm{rad}$)

$$\Delta V_\mathrm{oc,sub-gap}^\mathrm{rad} = V_\mathrm{oc}^\mathrm{SQ} - V_\mathrm{oc}^\mathrm{rad}. \qquad \mathbf{(S31)}$$

As demonstrated by **Figure S4**, regardless of the amount of sub-gap absorption that is present, the short-circuit current density is roughly the same. We therefore assume that $J_\mathrm{sc} \approx J_\mathrm{sc}^\mathrm{SQ} = q\int_{E_\mathrm{opt}}^{\infty} \Phi_\mathrm{source}(E)\,\mathrm{d}E$. The dark saturation current density, however, is strongly dependent on the absorption parameters. By combining Equation **(S31)** with Equation **(S3)**, one may write the deviation from $V_\mathrm{oc}^\mathrm{SQ}$ in the radiative limit ($\mathrm{EQE_{EL}} = 1$) as

$$\Delta V_\mathrm{oc,sub-gap}^\mathrm{rad} = \frac{kT}{q}\ln\left[\frac{J_0^\mathrm{rad}}{J_0^\mathrm{SQ}}\frac{J_0^\mathrm{SQ}+J_\mathrm{sc}^\mathrm{SQ}}{J_0^\mathrm{rad}+J_\mathrm{sc}}\right] \approx \frac{kT}{q}\ln\left[\frac{J_0^\mathrm{rad}}{J_0^\mathrm{SQ}}\frac{J_0^\mathrm{SQ}+J_\mathrm{sc}^\mathrm{SQ}}{J_0^\mathrm{rad}+J_\mathrm{sc}^\mathrm{SQ}}\right]. \qquad \mathbf{(S32)}$$

In the limit that $J_\mathrm{sc}^\mathrm{SQ} \gg J_0^\mathrm{SQ}$ and $J_\mathrm{sc}^\mathrm{SQ} \gg J_0^\mathrm{rad}$, which, as highlighted by **Figure S4**, can safely be assumed for $E_\mathrm{opt} > 1.0$ eV, Equation **(S32)** reduces to $\Delta V_\mathrm{oc,sub-gap}^\mathrm{rad} \approx \frac{kT}{q}\ln\left[\frac{J_0^\mathrm{rad}}{J_0^\mathrm{SQ}}\right]$. Substituting in the expressions for $J_0^\mathrm{rad}$ obtained above, approximations for the radiative open-circuit voltage loss induced by sub-gap absorption can be established for the different cases.

## S6. Power Conversion Efficiency Limits

By combining all the theory outlined in **Sections S1-S5**, the power conversion efficiency could be simulated in the radiative limit for the $\mathrm{EQE_{PV}}$ models based on Equation **(S21)** (for a variety of Urbach energies) and Equation **(S20)** (for a variety of $\sigma_S$) – the results are illustrated for the former in **Figure S5**, and for the latter in **Figure S6**. In **panel a** of those figures, the PCE is plotted as a function of the optical gap under the CIE LED-B4 spectrum at 500 lux. Whereas, in panel **b** of those figures, the maximum PCE is plotted as a function of the illuminance; the optical gaps that produce the maximum PCEs in the Urbach tail model are provided in **Table S2**, whereas the corresponding optical gaps for the $\mathrm{EQE_{PV}}$ model based on Equation **(S20)** are summarized in **Table S3**.

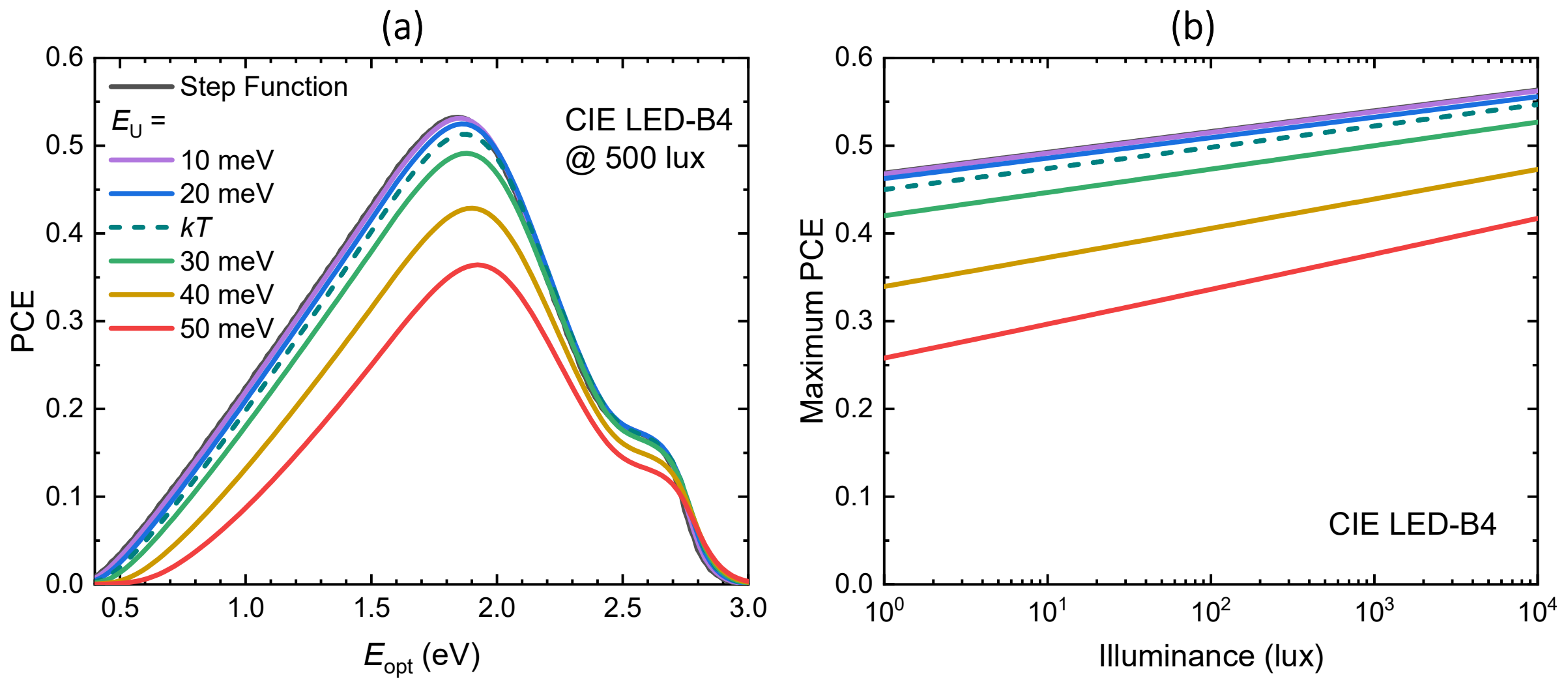

**Figure S5**: Power conversion efficiency in the radiative limit under CIE LED-B4 illumination for a variety of Urbach energies. **(a)** PCE versus optical gap at 500 lux. **(b)** Maximum PCE versus illuminance, with the highest-performing $E_{\mathrm{opt}}$ values summarised in **Table S2**. The $E_{\mathrm{U}} = 0$ case is equivalent to a step-function $\mathrm{EQE_{PV}}$. The $E_{\mathrm{U}} = kT$ case is indicated by the dashed line.

From **Figure S5**, one can see that sub-gap tails with Urbach energies less than 20 meV induce such little open-circuit voltage losses that the PCE in the radiative limit is essentially the same as the SQ model. For Urbach energies greater than 20 meV, however, the losses become more considerable. Consequently, the PCE drops by around 3% from the SQ model value when a sub-gap tail with $E_{\mathrm{U}} = kT$ is included (at room temperature, $T = 293.15$ K), as illustrated by the dashed line. Comparing these Urbach energy-dependent results with the energetic disorder-dependent results of **Figure S6**, one can see that a large $E_{\mathrm{U}}$ on the order of $50$ meV is far more detrimental to the PCE in the thermodynamic limit than a large energetic disorder on the order of $100$ meV. However, higher levels of energetic disorder still thermodynamically constrain the PCE more than lower levels of energetic disorder.

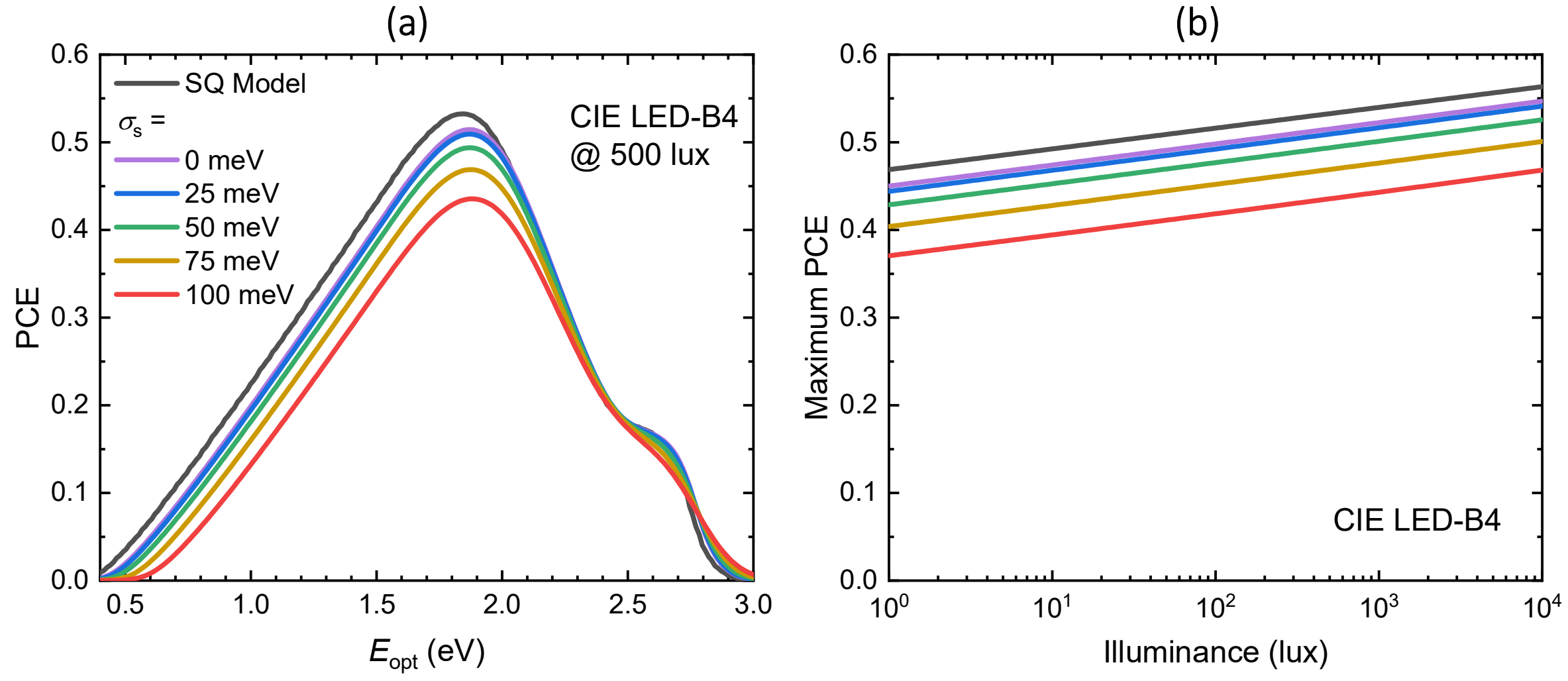

**Figure S6**: Power conversion efficiency in the radiative limit under CIE LED-B4 illumination for a variety of excitonic static disorder $\sigma_{\mathrm{s}}$. **(a)** PCE versus optical gap at 500 lux. **(b)** Maximum PCE versus illuminance, with the highest-performing $E_{\mathrm{opt}}$ values summarised in **Table S3**. The SQ model has also been illustrated by the black curves.

In **Figure 5** of the main text, the PCE of IPVs has been simulated under the CIE LED-B4 standard illuminant. These PCEs have also been simulated under illumination by the 'warm white' 2700K LED and 'cool white' 4000K LED spectra with the results being plotted in **Figure S7**. We note that similar results are obtained to the ones described in the main text for CIE LED-B4. The PCEs have also been simulated under three standard fluorescent spectra: CIE FL-2, CIE FL-7, and CIE FL-11. [4] The results are plotted in **Figure S8.**

**Table S2**: The optical gaps that produce the maximum PCEs illustrated in **Figure S5b**, provided alongside the Urbach energies that describe the curve.

| **Urbach Energy, $E_U$ (meV)** | **Best Optical Gap (eV)** |
|---|---|
| 0 (SQ Model) | 1.830 |
| 10 | 1.846 |
| 20 | 1.878 |
| $kT = 25.26$ | 1.878 |
| 30 | 1.878 |
| 40 | 1.910 |
| 50 | 1.910 |

**Table S3**: The optical gaps that produce the maximum PCEs illustrated in **Figure S6b**, provided alongside the excitonic static disorder values that describe the curve.

| **Excitonic Static Disorder, $\sigma_s$ (meV)** | **Best Optical Gap (eV)** |
|---|---|
| SQ Model | 1.830 |
| 0 | 1.878 |
| 25 | 1.870 |
| 50 | 1.872 |
| 75 | 1.875 |
| 100 | 1.881 |

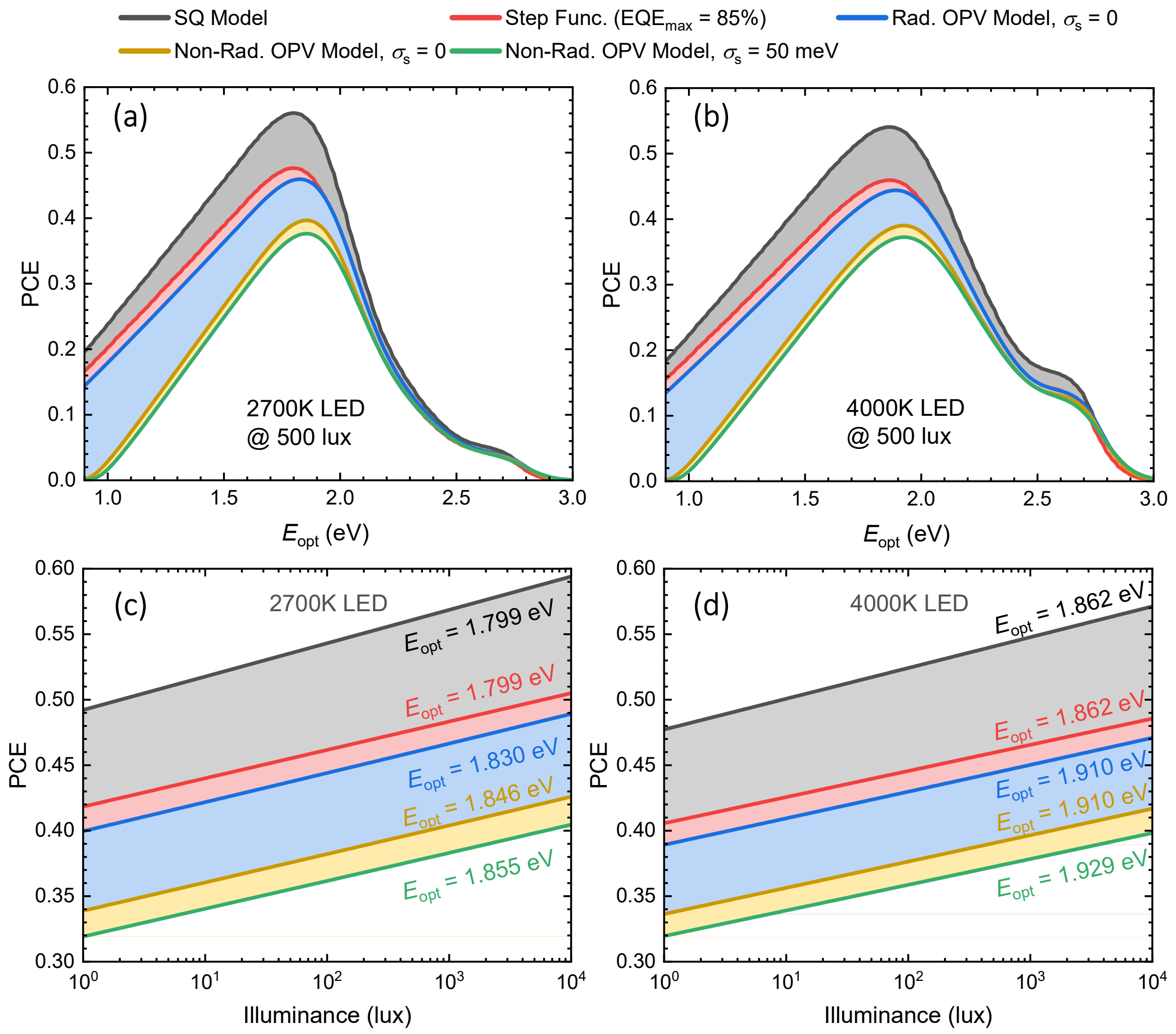

**Figure S7:** Power conversion efficiency in the SQ model (black), in the step function case determined using a step function with $\text{EQE}_{\text{max}} = 0.85$ (red), in the radiative and non-radiative OPV model for the $\sigma_s = 0$ meV case (blue and gold, respectively), and in the non-radiative, disordered OPV model with $\sigma_s = 50$ meV case (green). The latter is determined using a combination of the OPV model for the $\text{EQE}_{\text{PV}}$ and the realistic non-radiative open-circuit voltage model described in the main text. The grey shaded regions indicate loss due to non-unity $\text{EQE}_{\text{max}}$, while the red shaded regions indicate the loss induced by a sub-gap Urbach tail with $E_U = kT$. Finally, the blue and gold shaded regions indicate the PCE losses attributed to non-radiative open-circuit voltage loss, and energetic disorder, respectively. **(a)** and **(b)** show the PCE versus optical gap at an illuminance of 500 lux for the 2700K LED and the 4000K LED, respectively.

Whereas **(c)** and **(d)** show the PCE versus illuminance for the best-performing optical gap (inset with the curves), for the 2700K LED and 4000K LED, respectively.

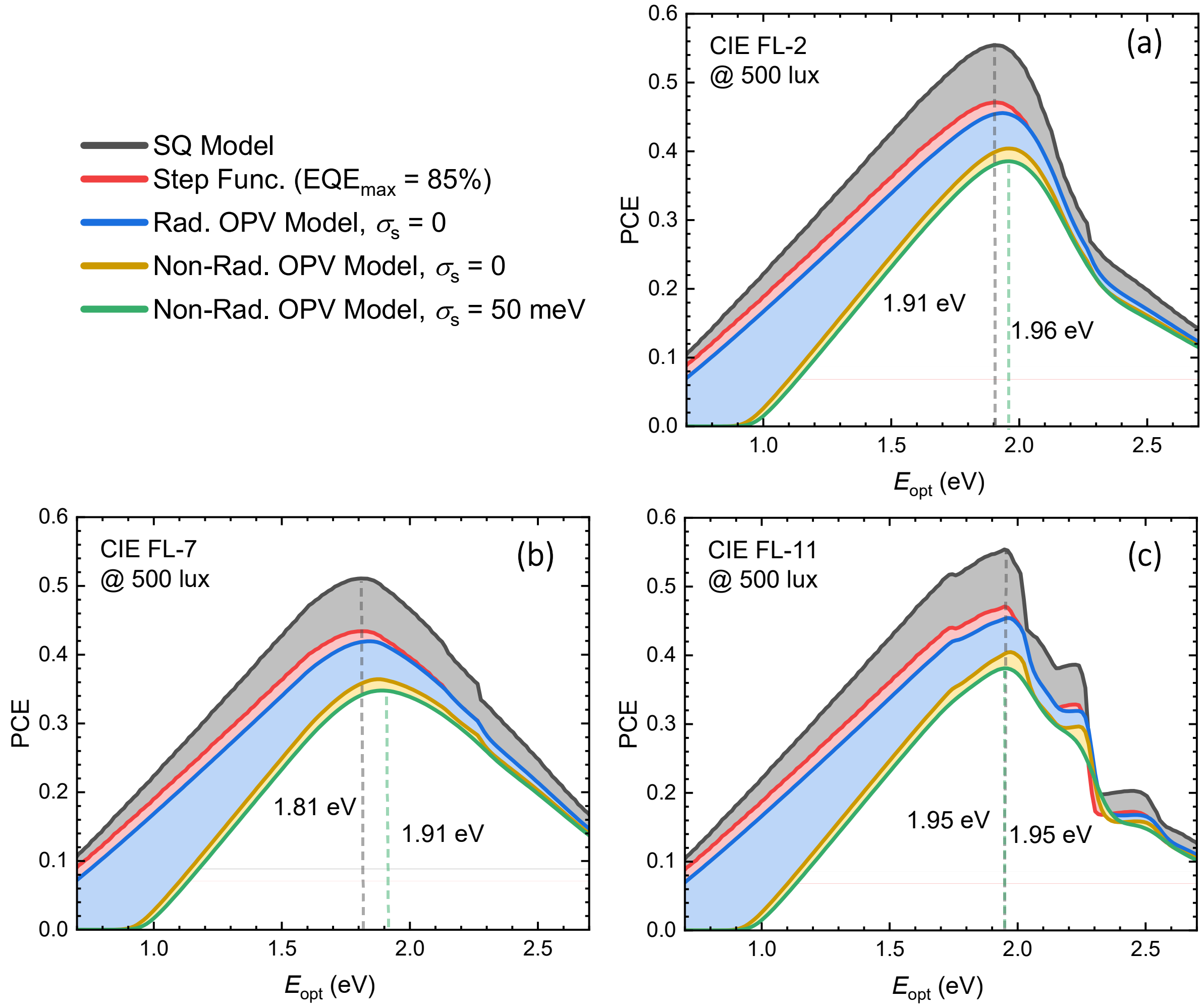

**Figure S8:** Power conversion efficiency versus optical gap under three standard CIE spectra: **(a)** FL-2, **(b)** FL-7, and **(c)** FL-11.[4] The black curves indicate the PCE in the SQ model, while the red curves indicate the PCE in the step function $EQE_{PV}$ model (with $EQE_{max} = 0.85$). The grey shaded region indicates the PCE loss induced by non-unity $EQE_{max}$. The blue and gold curves, on the other hand, indicate the PCE in the radiative and non-radiative OPV model (with $EQE_{max} = 0.85$ and $\sigma_s = 0$). The blue shaded region

indicates the PCE loss induced by a sub-gap Urbach tail with $E_{\mathrm{U}} = kT$. Finally, the green curve indicates the PCE in the non-radiative, disordered OPV model (same as gold curve with $\sigma_{\mathrm{s}} = 50$ meV), with the gold shaded region indicating the PCE loss induced by a disordered density of states.

# Part II – Experimental Results and Analysis

## S7. Material Definitions for Figure 6 of Main Text

**BQR**: N~2~-[7-(3,4-dimethoxyphenyl)quinoxalin-2-yl]-N-methylglycinamide

**BTP-eC9:** 2,2'-[[12,13-Bis(2-butyloctyl)-12,13-dihydro-3,9-dinonylbisthieno[2",3":4',5']thieno[2',3':4,5]pyrrolo[3,2-e:2',3'-g][2,1,3]benzothiadiazole-2,10-diyl]bis[methylidyne(5,6-chloro-3-oxo-1H-indene-2,1(3H)-diylidene) ]]bis[propanedinitrile]

**EH-IDTBR**: (5Z)-3-ethyl-2-sulfanylidene-5-[[4-[9,9,18,18-tetrakis(2-ethylhexyl)-15-[7-[(Z)-(3-ethyl-4-oxo-2-sulfanylidene1,3-thiazolidin-5-ylidene)methyl]-2,1,3-benzothiadiazol-4-yl]-5,14-dithiapentacyclo[10.6.0.03,10.04,8.013,17]octadeca-1(12),2,4(8),6,10,13(17),15-heptaen-6-yl]-2,1,3-benzothiadiazol-7-yl]methylidene]-1,3-thiazolidin-4-one

**ITIC:** 3,9-bis(2-methylene-(3-(1,1-dicyanomethylene)-indanone))-5,5,11,11-tetrakis(4-hexylphenyl)-dithieno[2,3-*d*:2′,3′-*d*′]-s-indaceno[1,2-*b*:5,6-*b*′]dithiophene

**O-IDTBR:** (5Z,5'Z)-5,5'-((7,7'-(4,4,9,9-tetraoctyl-4,9-dihydro-s-indaceno[1,2-b:5,6-b']dithiophene-2,7-diyl)bis(benzo[c][1,2,5]thiadiazole-7,4-diyl))bis(methanylylidene))bis(3-ethyl-2-thioxothiazolidin-4-one)

**PBDB-T:** Poly[(2,6-(4,8-bis(5-(2-ethylhexyl)thiophen-2-yl)-benzo[1,2-*b*:4,5-*b*′]dithiophene))-*alt*-(5,5-(1',3'-di-2-thienyl-5′,7′-bis(2-ethylhexyl)benzo[1′,2′-*c*:4′,5′-*c*′]dithiophene-4,8-dione)]

**$PC_{71}BM$:** [6,6]-phenyl-C71-butyric acid methyl ester

**PDINO:**2,9-bis[3-(dimethyloxidoamino)propyl]anthra[2,1,9-def:6,5,10-*d′e′f′*]diisoquinoline-1,3,8,10(2*H*,9*H*)-tetrone

**PEDOT:PSS:** Poly(3,4-ethylenedioxythiophene) polystyrene sulfonat

**PM6:** Poly[(2,6-(4,8-bis(5-(2-ethylhexyl-3-fluoro)thiophen-2-yl)-benzo[1,2-*b*:4,5-*b*′]dithiophene))-*alt*-(5,5-(1′,3′-di-2-thienyl-5′,7′-bis(2-ethylhexyl)benzo[1′,2′-*c*:4′,5′-*c*′]dithiophene-4,8-dione)]

**Y6:** 2,2′-((2*Z*,2′*Z*)-((12,13-bis(2-ethylhexyl)-3,9-diundecyl-12,13-dihydro-[1,2,5]thiadiazolo[3,4-*e*]thieno[2″,3″:4′,5′]thieno[2′,3′:4,5]pyrrolo[3,2-*g*]thieno[2′,3′:4,5]thieno[3,2-*b*]indole-2,10-diyl)bis(methanylylidene))bis(5,6-difluoro-3-oxo-2,3-dihydro-1*H*-indene-2,1-diylidene))dimalononitrile

**Crystalline Silicon:** Commercial crystalline silicon solar cell (Part number: KXOB22-12X1)

**Amorphous Silicon:** a-Si:H thin-film solar cell made by Trony (SC80125s-8)

## S8. Methodology for Estimating Indoor Performance Using One-Sun Measurements

To make realistic predictions for the figures-of-merit of photovoltaic materials in indoor settings, we devised a methodology that takes experimentally-determined measurements of the photovoltaic external quantum efficiency and the open-circuit voltage under one-Sun conditions ($V_{\mathrm{oc}}^{\odot}$) as inputs. As illustrated by the left-hand 'AM1.5 Global Conditions' pane of the block diagram in **Figure S9**, $V_{\mathrm{oc}}^{\odot}$ can be determined from a device's current-voltage curve. The $\mathrm{EQE_{PV}}$ spectrum, on the other hand, can be used to determine the short-circuit and dark saturations current densities using

$$J_{\mathrm{sc}} = q \int_{E_{\mathrm{lower}}}^{\infty} \mathrm{EQE_{PV}}(E) \Phi_{\mathrm{source}}(E) \, \mathrm{d}E \tag{S33a}$$

$$J_0^{\mathrm{rad}} = q \int_{E_{\mathrm{lower}}}^{\infty} \mathrm{EQE_{PV}}(E) \Phi_{\mathrm{bb}}(E) \, \mathrm{d}E. \tag{S33b}$$

Here, $E_{\mathrm{lower}}$ is the lower limit of the integral, which must be varied to determine the true radiative open-circuit voltage $V_{\mathrm{oc}}^{\mathrm{rad}}$, as described in the work of Zarrabi et al.. [23] It must be stressed that $V_{\mathrm{oc}}^{\mathrm{rad}}$ is *not* a spectral quantity; varying the lower limit of the integral just makes it appear that way. The true $V_{\mathrm{oc}}^{\mathrm{rad}}$ in the thermodynamic limit can be extracted from plots like the right-hand graph in the left pane of **Figure S9** by identifying the point where a plateau is reached. Using this $V_{\mathrm{oc}}^{\mathrm{rad}}$, the non-radiative open-circuit voltage loss under one-Sun conditions can be extracted using $\Delta V_{\mathrm{oc}}^{\mathrm{nr}} = V_{\mathrm{oc}}^{\mathrm{rad}} - V_{\mathrm{oc}}^{\odot}$. Assuming this $\Delta V_{\mathrm{oc}}^{\mathrm{nr}}$ is the minimum open-circuit voltage experienced by the device under any set of illumination conditions, then the figures-of-merit can be predicted in indoor conditions using the approach outlined in **Section S2**, as illustrated in the left-hand pane of **Figure S9**.

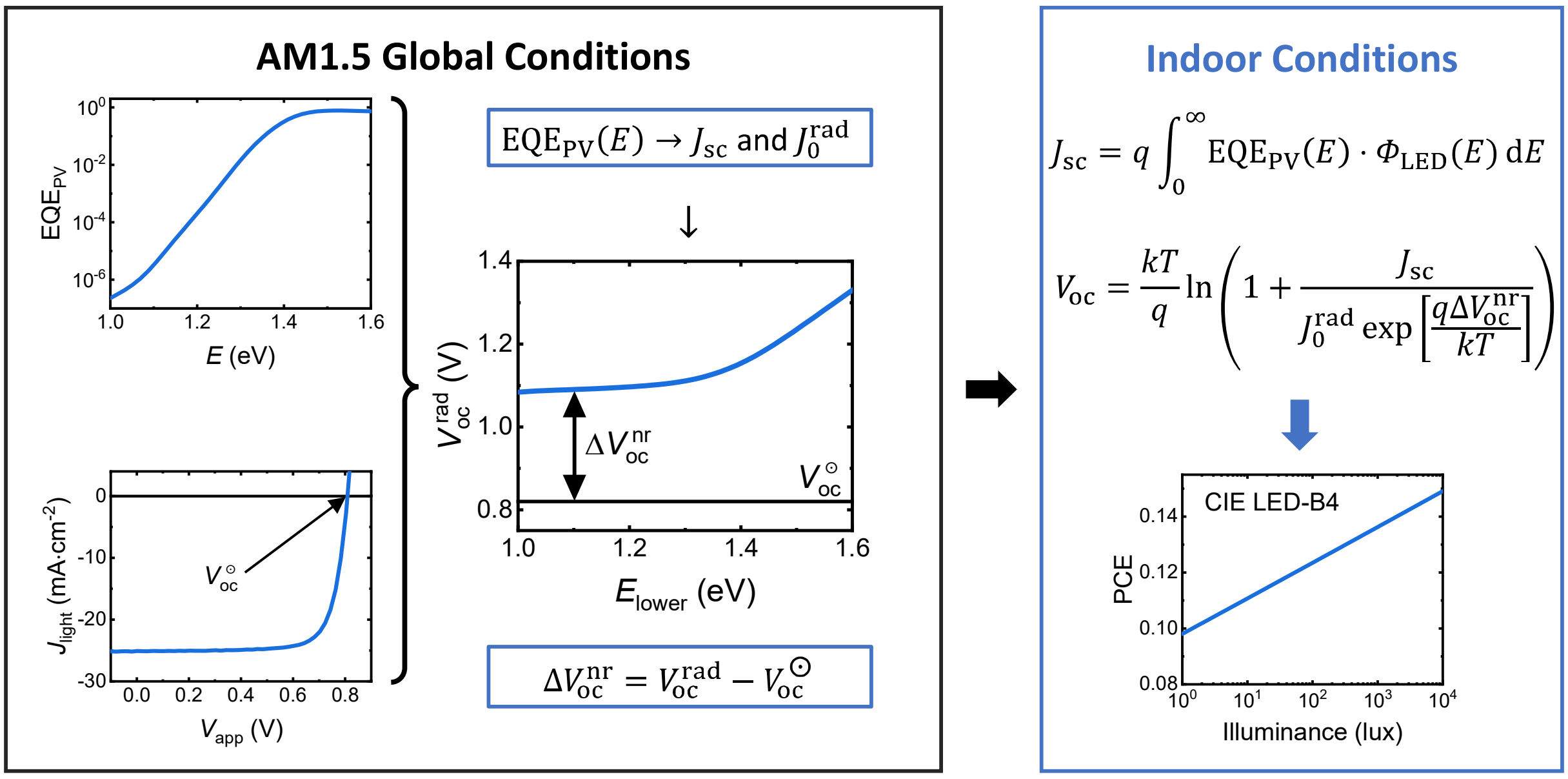

**Figure S9**: A block diagram illustrating how one-Sun (AM1.5 global) measurements can be used to predict device performance under indoor conditions. The left-hand pane shows how a photovoltaic external quantum efficiency spectrum and a current-density curve can be used, in combination, to extract the true $V_{oc}^{rad}$ which, in turn, can be used to evaluate the non-radiative open-circuit voltage loss, $\Delta V_{oc}^{nr}$. This loss is then carried across to the right-hand pane, illustrating the calculation of figures-of-merit under arbitrary indoor conditions.

## S9. Extracting Parameters from Photovoltaic External Quantum Efficiency Spectra

For the OPV systems re-contextualized from previous works to predict indoor peformance in this work, the optical gap and energetic disorder were determined using a methodology outlined in prior work by the authors.[17] In that work, photovoltaic external quantum efficiency spectrum were fit with

$$\mathrm{EQE_{PV}}(E) = \mathrm{EQE_0}[1 - \exp(-\alpha(E)d)], \tag{S34}$$

where $\mathrm{EQE_0}$ is a pre-factor, $d$ is the active layer thickness, and $\alpha$ is the exciton absorption coefficient. In this work, however, we make use of the weak $\alpha d$ limit (valid well below the gap) to write $\mathrm{EQE_{PV}}(E)$ as seen in Equation **(S20)**.[24] In this limit, the apparent Urbach energy ($E_\mathrm{U}^\mathrm{app}$), defined by[16, 25]

$$E_\mathrm{U}^\mathrm{app}(E) = \left[\frac{\partial \ln\left(\mathrm{EQE_{PV}}(\epsilon)\right)}{\partial \epsilon}\right]^{-1}\Bigg|_{\epsilon=E}, \tag{S35}$$

may be approximated with

$$E_\mathrm{U}^\mathrm{app}(E) \approx kT\left[1 + \frac{1 + \mathrm{erf}\left(\frac{E - E_\mathrm{opt}}{\sigma_\mathrm{s}\sqrt{2}}\right)}{\exp\left(\frac{E - E_\mathrm{opt} + \frac{\sigma_\mathrm{s}^2}{2kT}}{kT}\right)\mathrm{erfc}\left(\frac{E - E_\mathrm{opt} + \frac{\sigma_\mathrm{s}^2}{kT}}{\sigma_\mathrm{s}\sqrt{2}}\right)}\right]. \tag{S36}$$

Note that $E_\mathrm{opt} \gg \sigma_\mathrm{s}$ has been assumed. The spectral behavior of Equation **(S36)** is plotted normalized to the thermal energy $kT$ for varied $E_\mathrm{opt}$ in **Figure S10a**, and for varied $\sigma_\mathrm{s}$ in **Figure S10b**. From these graphs, it can be seen that well below the gap, $E_\mathrm{U}^\mathrm{app} \to kT$, regardless of the $\sigma_\mathrm{s}$ value. Furthermore, as $E \to E_\mathrm{opt}$, the apparent Urbach energy tends to infinity; a shift in $E_\mathrm{opt}$ corresponds with an equivalent shift in the $E_\mathrm{U}^\mathrm{app}$ spectrum. Whereas a change in $\sigma_\mathrm{s}$ on the other hand corresponds with a broadened transition between the $E_\mathrm{U}^\mathrm{app} = kT$ and the $E_\mathrm{U}^\mathrm{app} \to \infty$ regimes.

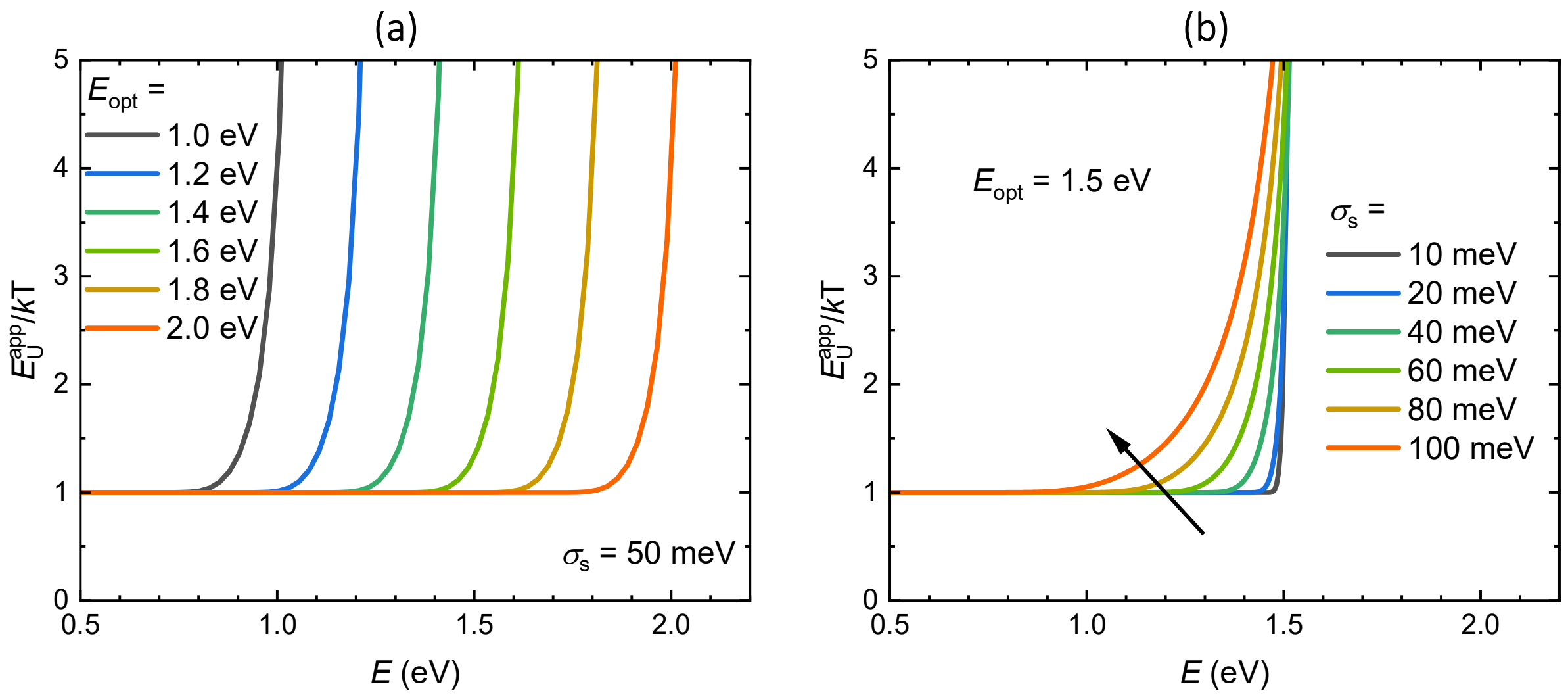

**Figure S10**: The apparent Urbach energy approximation given by Equation **(S36)**, plotted normalized to the thermal energy $kT$ as a function of the photon energy $E$, for varied $E_{\mathrm{opt}}$ in (**a**) and varied $\sigma_{\mathrm{s}}$ in (**b**).

Using Equation **(S36)**, the optical gap and energetic disorder of several technologically-relevant donor:acceptor OPV blends were estimated using their apparent Urbach spectra, which were determined from the corresponding reported $\mathrm{EQE_{PV}}$ spectra using Equation **(S35)**. The $\mathrm{EQE_{PV}}$ spectra themselves were also fit with the methodology outlined in our previous work. [17] The extracted values are inserted into the respective graphs throughout the remainder of this section, with the values extracted using Equation **(S36)** likely to be more accurate due to one of the free parameters, $\mathrm{EQE_{max}}$, being removed from the fitting. The spectra are plotted in **Figure S11** to **Figure S19**; the extracted values are summarized in **Table S4**.

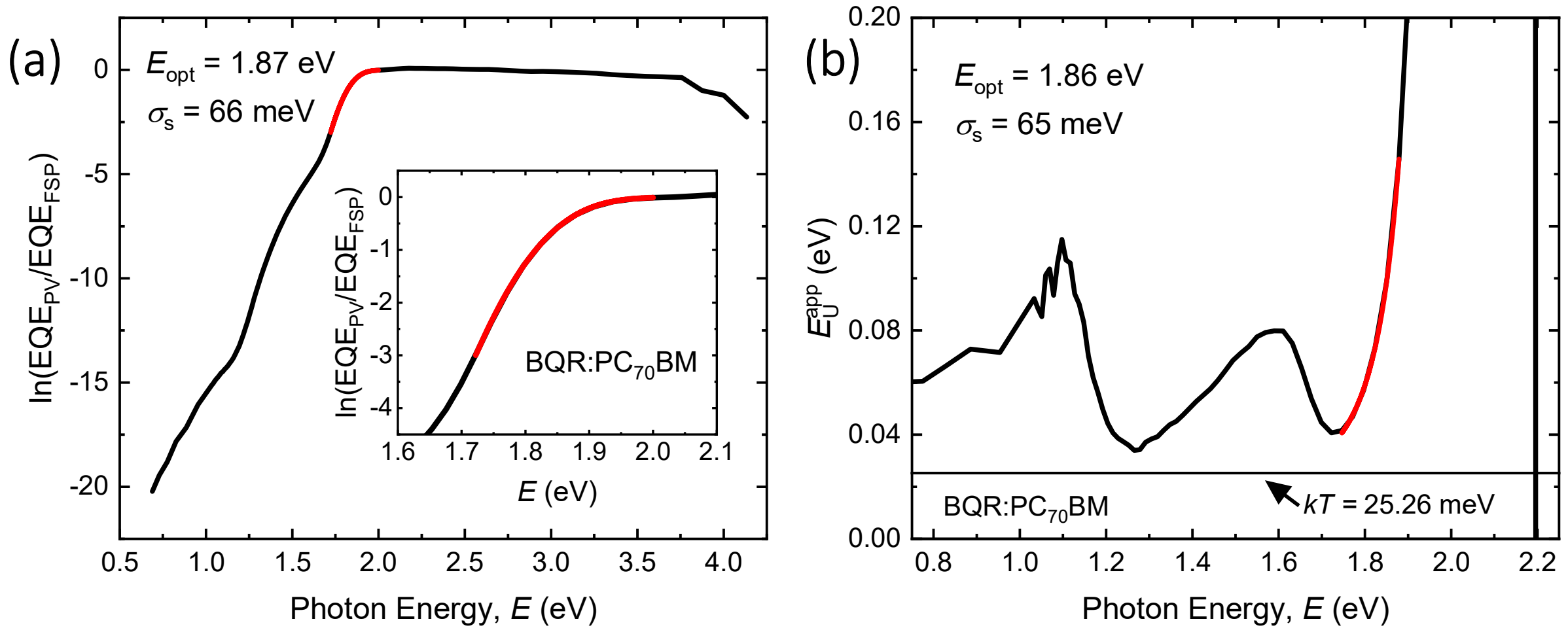

**Figure S11**: Experimentally-determined and simulated parameters for BQR:$PC_{70}BM$ (**a**) The experimentally-determined photovoltaic external quantum efficiency spectrum, normalized to its value at the first saturation peak ($EQE_{FSP}$) and (**b**) its corresponding apparent Urbach energy spectrum. Both spectra are fit with their respective models – the extracted values for the narrower optical gap component ($PC_{70}BM$) are included.

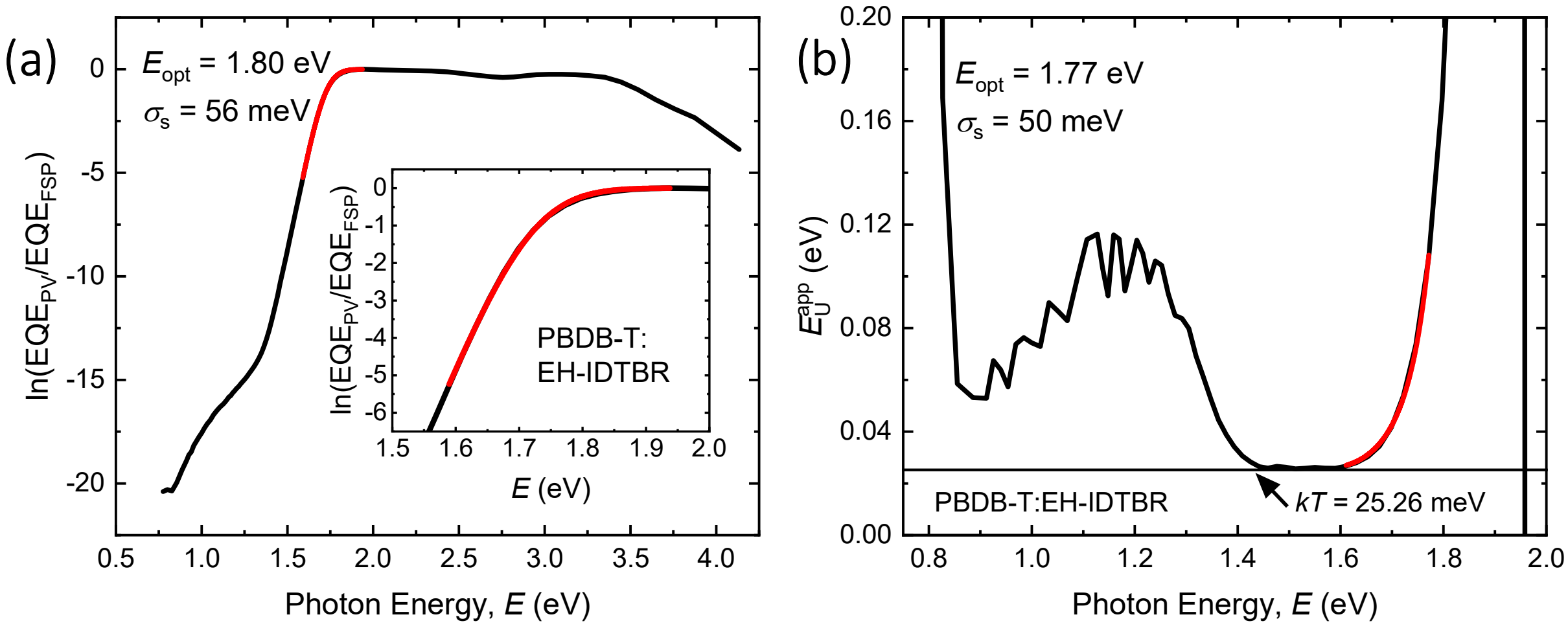

**Figure S12**: (**a**) The photovoltaic external quantum efficiency spectrum for PBDB-T:EH-IDTBR, normalized to its value at the first saturation peak ($EQE_{FSP}$) and (**b**) its corresponding apparent Urbach energy spectrum. As illustrated by the red curves, both spectra are fit with their respective models – $EQE_{PV}$

with the methodology outlined in our previous work [17] and $E_U^{app}$ with Equation **(S36)** – and the extracted values for the narrower optical gap component (EH-IDTBR) are included.

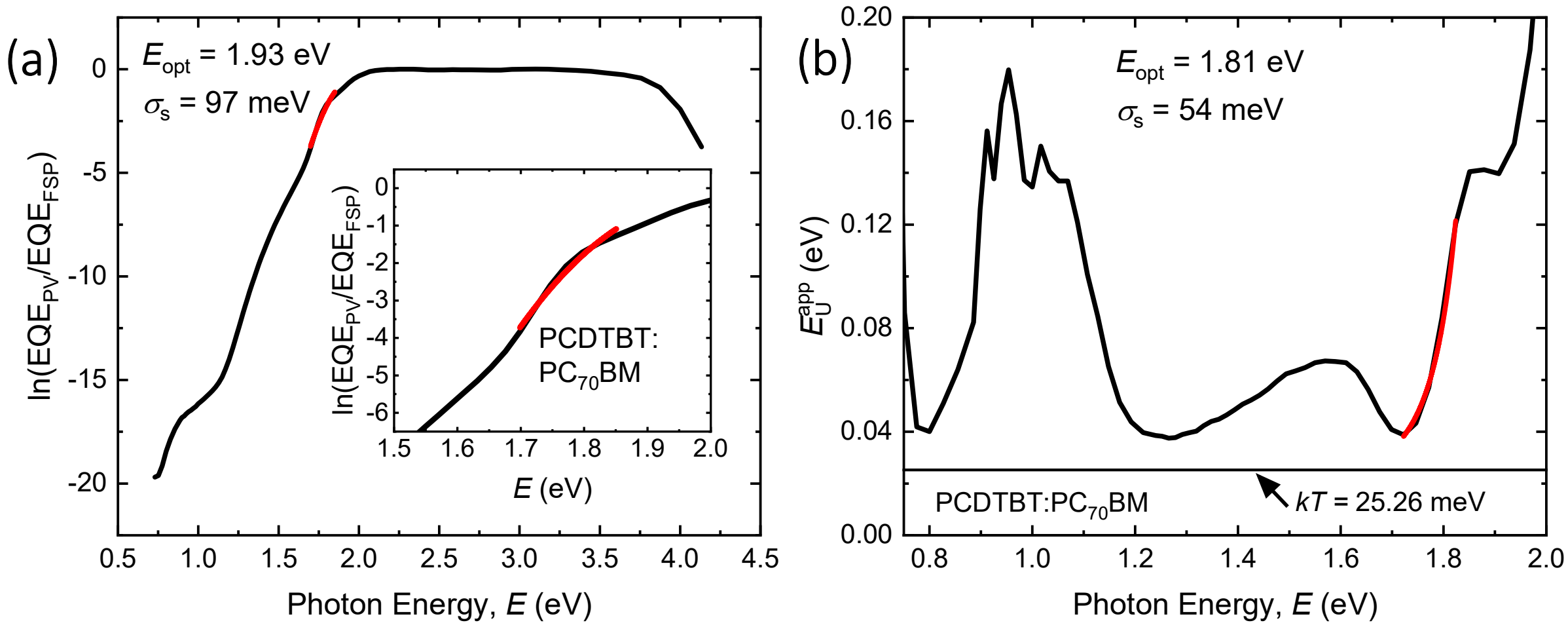

**Figure S13**: (**a**) The photovoltaic external quantum efficiency spectrum for PCDTBT:$PC_{70}BM$, normalized to its value at the first saturation peak ($EQE_{FSP}$) and (**b**) its corresponding apparent Urbach energy spectrum. As illustrated by the red curves, both spectra are fit with their respective models – $EQE_{PV}$ with the methodology outlined in our previous work [17] and $E_U^{app}$ with Equation **(S36)** – and the extracted values for the narrower optical gap component (PCDTBT) are included.

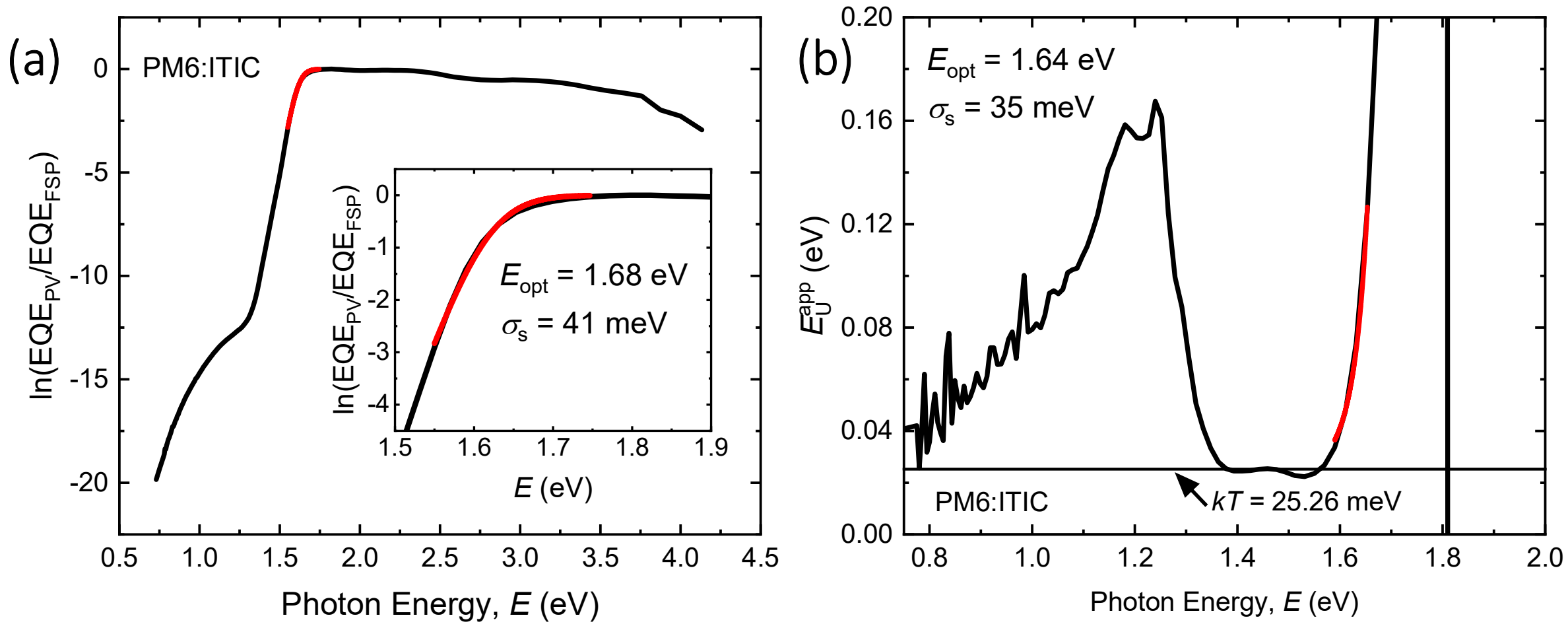

**Figure S14**: (**a**) The photovoltaic external quantum efficiency spectrum for PM6:ITIC, normalized to its value at the first saturation peak ($EQE_{FSP}$) and (**b**) its corresponding apparent Urbach energy spectrum. As illustrated by the red curves, both spectra are fit with their respective models – $EQE_{PV}$ with the methodology outlined in our previous work [17] and $E_U^{app}$ with Equation **(S36)** – and the extracted values for the narrower optical gap component (ITIC) are included.

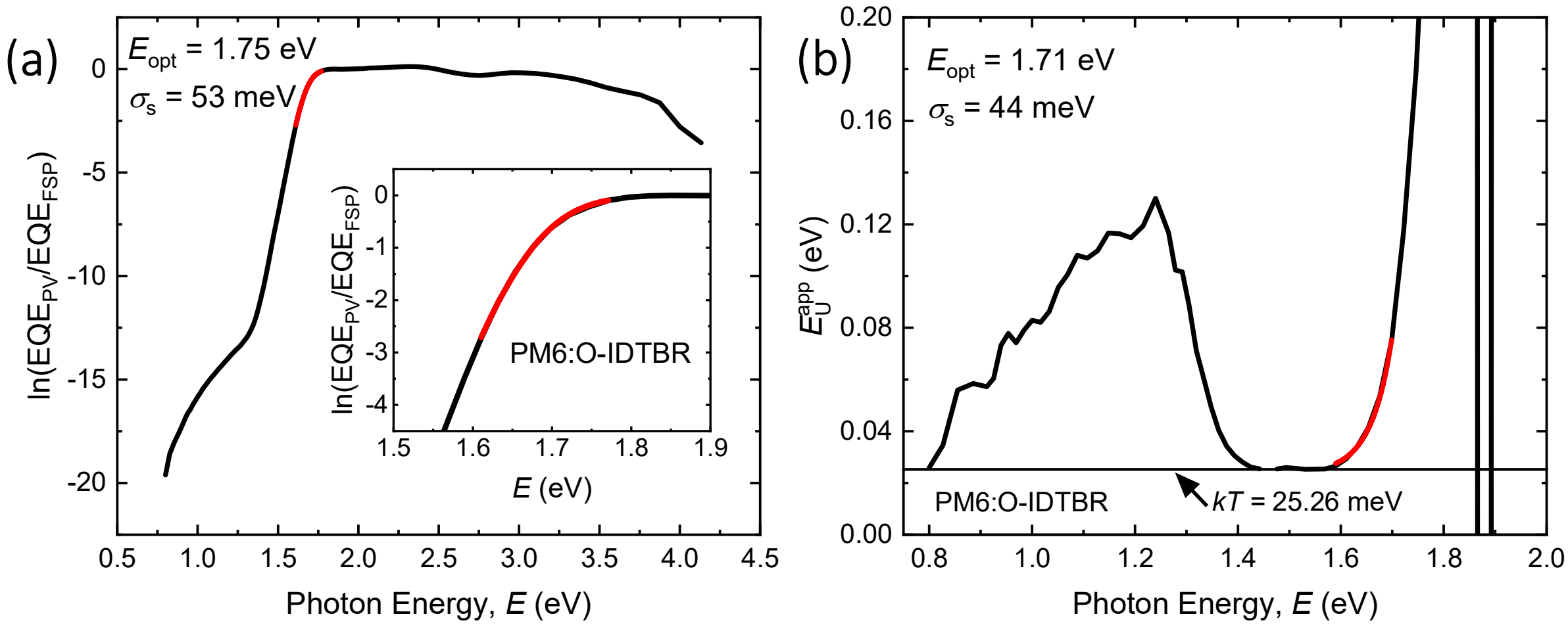

**Figure S15**: (**a**) The photovoltaic external quantum efficiency spectrum for PM6:O-IDTBR, normalized to its value at the first saturation peak ($EQE_{FSP}$) and (**b**) its corresponding apparent Urbach energy spectrum. As illustrated by the red curves, both spectra are fit with their respective models – $EQE_{PV}$ with the methodology outlined in our previous work [17] and $E_U^{app}$ with Equation **(S36)** – and the extracted values for the narrower optical gap component (O-IDTBR) are included.

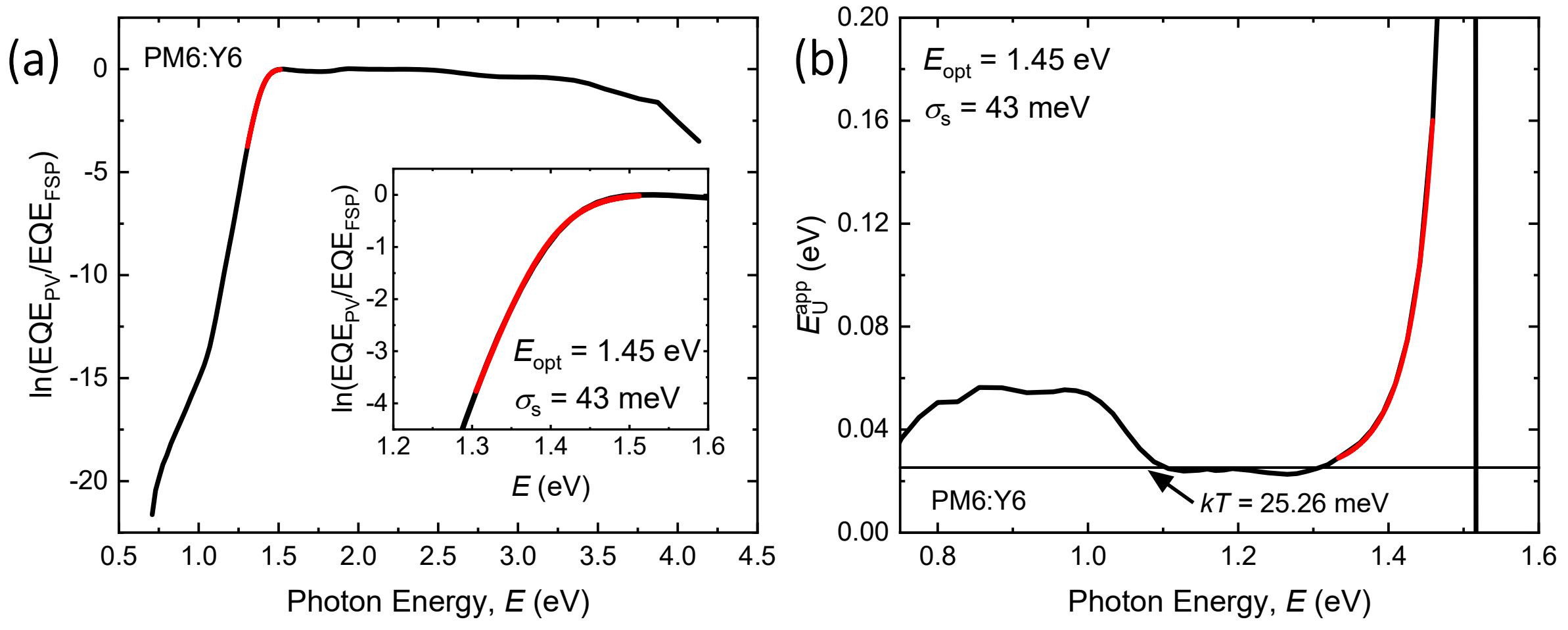

**Figure S16**: (**a**) The photovoltaic external quantum efficiency spectrum for PM6:Y6, normalized to its value at the first saturation peak ($EQE_{FSP}$) and (**b**) its corresponding apparent Urbach energy spectrum. As illustrated by the red curves, both spectra are fit with their respective models – $EQE_{PV}$ with the methodology outlined in our previous work [17] and $E_U^{app}$ with Equation **(S36)** – and the extracted values for the narrower optical gap component (Y6) are included.

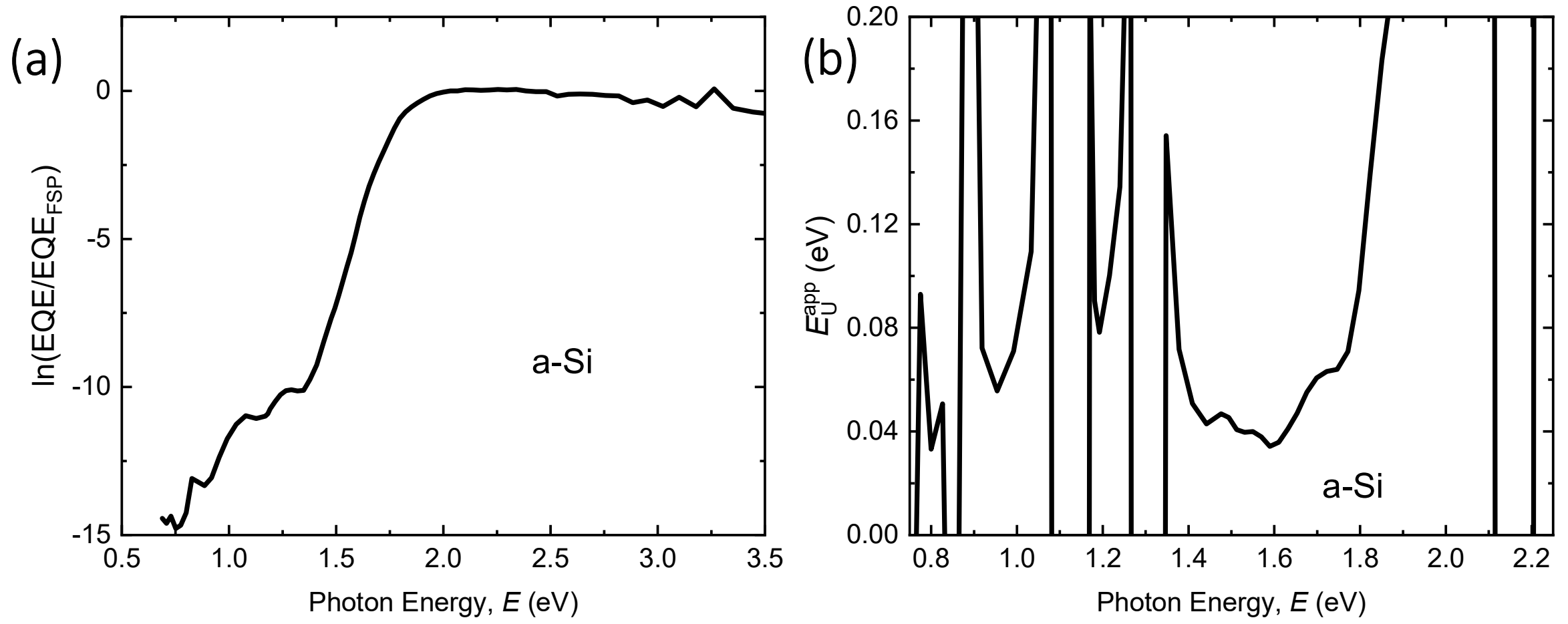

**Figure S17: (a)** The photovoltaic external quantum efficiency spectrum normalized to its first saturation peak value ($EQE_{FSP}$) and **(b)** the resultant apparent Urbach energy spectrum for amorphous silicon (a-Si). While these spectra were used to estimate the indoor performance, the optical gap was referenced from the literature. [26]

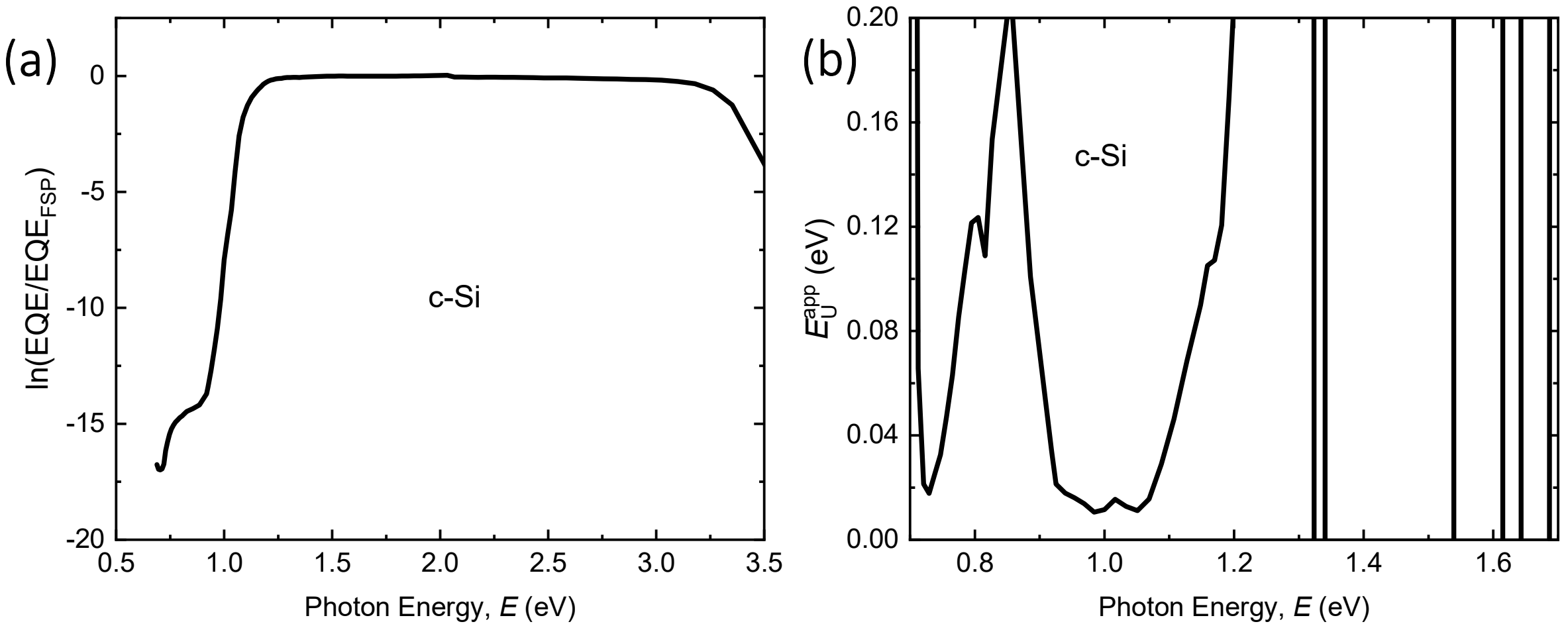

**Figure S18: (a)** The photovoltaic external quantum efficiency spectrum normalized to its first saturation peak value ($\mathrm{EQE_{FSP}}$) and **(b)** the resultant apparent Urbach energy spectrum for crystalline silicon (c-Si). While these spectra were used to estimate the indoor performance, the optical gap was referenced from the literature. [26]

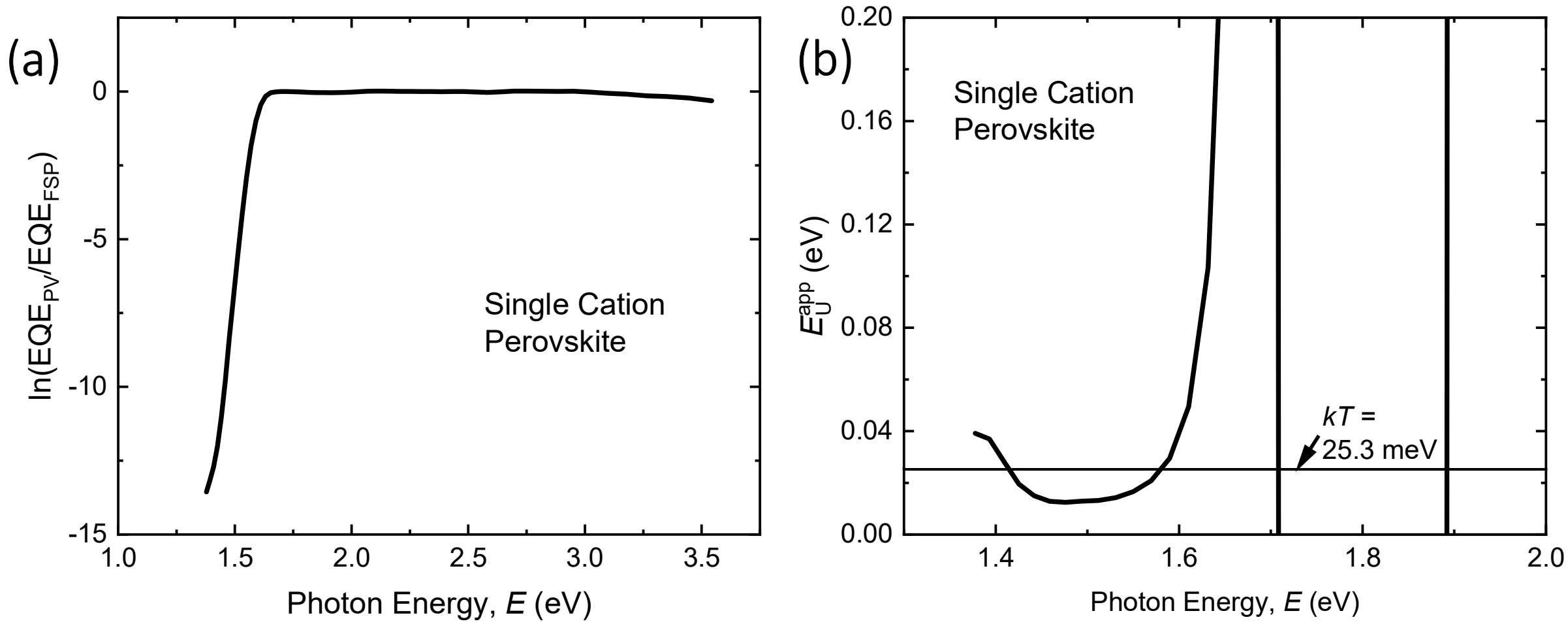

**Figure S19: (a)** The photovoltaic external quantum efficiency spectrum normalized to its first saturation peak value ($\mathrm{EQE_{FSP}}$) and **(b)** the resultant apparent Urbach energy spectrum for a single cation perovskite device. [20] This spectrum was used to estimate the indoor performance of the device; its optical gap is around 1.61 eV, [27] and its Urbach was previously found to be $E_\mathrm{U} = 13.0$ meV (around half the thermal energy, $kT$). [20, 27]

**Table S4**: Optical gap ($E_{\mathrm{opt}}$) and excitonic static disorder ($\sigma_{\mathrm{s}}$) values extracted from apparent Urbach energy spectra of **Figure S11-S19**. For the OPV systems, the values correspond to the narrower optical gap components of each respective blend, which have been underlined in the first column for clarity.

| System | $E_{\mathrm{opt}}$ (eV) | $\sigma_{\mathrm{s}}$ (meV)[a] |
|---|---|---|
| Crystalline Silicon | 1.12 [6] | - |
| Amorphous Silicon | 1.60 [26] | - |
| Single Cation Perovskite | 1.61 [27] | - |
| PM6:<u>Y6</u> | 1.44 | 43 |
| PBDB-T:<u>EH-IDTBR</u> | 1.77 | 50 |
| PM6:<u>BTP-eC9</u> | 1.42 | 43 |
| PM6:<u>ITIC</u> | 1.64 | 35 |
| PM6:<u>O-IDTBR</u> | 1.71 | 44 |
| BQR:<u>$PC_{70}BM$</u> | 1.86 | 65 |
| <u>PCDTBT</u>:$PC_{70}BM$ | 1.81 | 54 |

a – Excitonic static disorder values quoted estimated only for organic systems

## S10. Literature Data

To create Figure 6 in the main text, two sets of literature data were compiled. The first of these contained the photovoltaic figures-of-merit for the best-performing OPV systems in the literature – these are outlined in **Table S5**. The second set of data contained the experimentally-determined PCE versus optical gap for a variety of systems – these are outlined in **Table S6**.

**Table S5:** Optical gaps, open-circuit voltages under one Sun, and estimated non-radiative open-circuit voltage loss for the best-performing OPV systems in the literature (as compiled by Almora et al. [28, 29]). Also included are crystalline and amorphous silicon and a single cation perovskite. The non-radiative open-circuit voltage loss was determined using the technique outlined in Section S9. [23] The database label is used by the computational tool to map a particular system's open-circuit voltage to its $\mathrm{EQE_{PV}}$ spectrum.

| **Database Label** | $E_{opt}$ (eV) | Active Material | $V_{oc}^{\odot}$ (V) | Ref. | $E_{lower}$ (eV) | Estimated $\Delta V_{oc}^{nr}$ (V) |
|---|---|---|---|---|---|---|
| ORG_1 | 1.22 | BTB7-Th:ATT-9 | 0.663 | [30] | 1.124 | 0.275 |
| ORG_3 | 1.32 | PTB7-Th:IEICO-4F | 0.69 | [31] | 1.345 | 0.404 |
| ORG_5 | 1.34 | PTB7-Th:IEICO-4F | 0.712 | [32] | 1.253 | 0.335 |
| ORG_6 | 1.35 | PM6:mBzS-4F | 0.804 | [33] | 1.268 | 0.254 |
| ORG_7 | 1.35 | PM6:Y6 | 0.82 | [34] | 1.185 | 0.256 |
| ORG_8 | 1.36 | PM6:Y11 | 0.846 | [35] | 1.254 | 0.219 |
| ORG_9 | 1.37 | PM6:BTP-eC9:PC71BM | 0.856 | [36] | 1.267 | 0.221 |
| ORG_10 | 1.38 | PM6:BTP-T-3Cl:BTP-4Cl-BO | 0.857 | [37] | 1.225 | 0.231 |
| ORG_13 | 1.38 | PM6:BTP-eC9:L8-BO-F | 0.847 | [38] | 1.251 | 0.235 |
| ORG_14 | 1.38 | PM6:BTP-eC9:BTP-S9 | 0.862 | [39] | 1.262 | 0.223 |
| ORG_15 | 1.39 | PM6:Y6-1O:BO-4Cl | 0.848 | [40] | 1.264 | 0.244 |
| ORG_16 | 1.39 | D18:Y6 | 0.859 | [41] | 1.275 | 0.236 |
| ORG_18 | 1.39 | PB2:PBDB-TF:BTP-eC9 | 0.858 | [42] | 1.285 | 0.236 |
| ORG_19 | 1.39 | PM6:BTP-eC11:BTP-S2 | 0.872 | [43] | 1.253 | 0.223 |
| ORG_20 | 1.39 | PM6:BTP-eC9:ZY-4Cl | 0.863 | [44] | 1.278 | 0.229 |
| ORG_22 | 1.40 | PBDB-TF:L8-BO:BTP-eC9 | 0.869 | [45] | 1.274 | 0.231 |
| ORG_24 | 1.40 | PM6:AC9 | 0.867 | [46] | 1.249 | 0.225 |
| ORG_25 | 1.40 | PM6:CNS-6-8:Y6:PC71BM | 0.868 | [47] | 1.269 | 0.231 |
| ORG_27 | 1.40 | PTzBI-dF:BTP-TBr | 0.845 | [48] | 1.254 | 0.239 |
| ORG_28 | 1.41 | D18-Cl:PM6:Y6 | 0.871 | [49] | 1.283 | 0.232 |
| ORG_29 | 1.41 | PM6:PB2F:BTP-eC9 | 0.86 | [50] | 1.269 | 0.231 |
| ORG_31 | 1.41 | PBQx-TF:eC9-2Cl:F-BTA3 | 0.878 | [51] | 1.271 | 0.226 |
| ORG_32 | 1.42 | PBDB-TF:BTP-4F | 0.834 | [52] | 1.265 | 0.256 |
| ORG_34 | 1.43 | PBDB-T-2F:BTP-4F-P2EH | 0.88 | [53] | 1.316 | 0.252 |
| ORG_35 | 1.43 | PM6:IDST-4F | 0.82 | [54] | 1.304 | 0.314 |
| ORG_36 | 1.44 | PM6:PY-IT:BN-T | 0.955 | [55] | 1.363 | 0.207 |
| ORG_38 | 1.44 | PM6:L8-BO | 0.883 | [56] | 1.347 | 0.268 |
| ORG_40 | 1.45 | D18:L8-BO | 0.918 | [57] | 1.328 | 0.230 |
| ORG_41 | 1.45 | PM6:D18:L8-BO | 0.891 | [58] | 1.328 | 0.260 |
| ORG_43 | 1.46 | PM6:PY-DT | 0.949 | [59] | 1.363 | 0.227 |
| ORG_44 | 1.47 | PBDB-T-2Cl:BP-4F:MF1 | 0.882 | [60] | 1.406 | 0.303 |
| ORG_45 | 1.48 | PBDB-T:IDT-EDOT:PC71BM | 0.88 | [61] | 1.369 | 0.292 |
| ORG_46 | 1.50 | PM6:DTTC-4Cl | 0.92 | [62] | 1.472 | 0.374 |
| ORG_47 | 1.51 | PM6:SeTlC4Cl-DIO | 0.78 | [63] | 1.392 | 0.427 |
| ORG_48 | 1.52 | PBDB-T:IDT-EDOT:PC71BM | 0.85 | [61] | 1.397 | 0.355 |
| ORG_49 | 1.53 | PM6:SeTlC4Cl | 0.85 | [63] | 1.472 | 0.391 |
| ORG_50 | 1.54 | BTR:NITI:PC71BM | 0.94 | [64] | 1.466 | 0.314 |
| ORG_51 | 1.55 | PM6:IT-4F | 0.84 | [65] | 1.470 | 0.416 |
| ORG_52 | 1.56 | PBDB-T-2F:IT-4F | 0.826 | [66] | 1.443 | 0.424 |
| ORG_53 | 1.58 | PM6:DTTC-4F | 0.95 | [62] | 1.435 | 0.306 |
| ORG_54 | 1.58 | PBDB-T-SF:IT-4F | 0.88 | | 1.476 | 0.392 |
| ORG_55 | 1.61 | PM6:DTC-4F | 0.94 | [62] | 1.474 | 0.349 |
| ORG_56 | 1.61 | PBDB-T-2Cl:MF1 | 0.916 | [60] | 1.515 | 0.399 |
| ORG_58 | 1.62 | PTQ10:IDTPC | 0.93 | [68] | 1.490 | 0.380 |
| ORG_59 | 1.63 | PTQ10:IDIC-2F | 0.91 | [69] | 1.553 | 0.421 |
| ORG_60 | 1.64 | PTQ10:IDIC | 0.96 | [69] | 1.557 | 0.378 |
| ORG_61 | 1.65 | J51:ITIC | 0.82 | [70] | 1.535 | 0.513 |
| ORG_65 | 1.68 | PBDTTT-EFT:EHIDTBR | 1.03 | [71] | 1.571 | 0.323 |
| ORG_66 | 1.69 | PBT1-C:NFA | 0.878 | [72] | 1.594 | 0.507 |
| ORG_69 | 1.78 | PPDT2FBT:PC70BM | 0.786 | [73] | 1.686 | 0.617 |
| ORG_70 | 1.79 | BDT-ffBX-DT:PDI4 | 1.14 | [74] | 1.633 | 0.319 |
| ORG_71 | 1.79 | BDT-ffBX-DT:SFPDI | 1.23 | [74] | 1.655 | 0.232 |
| ORG_72 | 1.85 | BTR:PC71BM | 0.90 | [64] | 1.699 | 0.598 |
| ORG_73 | 1.85 | PBDB-T:PC71BM | 0.83 | [61] | 1.581 | 0.593 |
| ORG_74 | 1.86 | PBDB-T:NDP-Se-DIO | 0.94 | [75] | 1.749 | 0.595 |
| ORG_75 | 1.88 | PBDB-T-2Cl:PC61BM | 0.95 | [76] | 1.658 | 0.524 |
| ORG_76 | 1.93 | P3HT:TCBD14 | 0.79 | [77] | 1.848 | 0.824 |
| ORG_77 | 2.01 | P3HT:PCBM | 0.592 | [78] | 1.713 | 0.913 |
| ORG_78 | 1.45 | PM6:Y6 | 0.82 | [23] | 1.127 | 0.272 |
| ORG_79 | 1.81 | PCDTBT:PCBM | 0.90 | [23] | 1.216 | 0.361 |
| ORG_80 | 1.45 | PM6:ITIC | 0.98 | [23] | 1.409 | 0.321 |
| ORG_81 | 1.39 | PM6:BTP-eC9 | 0.84 | [79] | 1.240 | 0.249 |
| ORG_82 | 1.39 | PBDB-T:EH-IDTBR | 0.96 | [79] | 1.476 | 0.430 |
| ORG_83 | 1.86 | BQR:PCBM | 0.91 | [79] | 1.393 | 0.406 |
| ORG_84 | 1.71 | PM6:O-IDTBR | 1.07 | [79] | 1.378 | 0.276 |
| INO_1 | 1.17 | c-Si | 0.66 | [23] | 1.240 | 0.334 |
| INO_2 | 1.61 | a-Si:H | 0.90 | [16] | 1.442 | 0.460 |
| INO_3 | 1.27 | a-Si | 0.63 | [80] | 1.244 | 0.389 |
| PER_5759 | 1.61 | Single Cation Perovskite | 1.10 | [20] | 1.476 | 0.211 |

**Table S6:** Experimentally-determined photovoltaic figures-of-merit for a variety of OPV devices, under a selection of LED sources at varied light intensities. Where the type of LED source used was not specified, the entry in the Spectrum column is 'LED'.

| System | $E_{opt}$ | Spectrum | Illuminance | $V_{oc}^{\odot}$ | $J_{sc}$ | FF | PCE | Ref. |
|---|---|---|---|---|---|---|---|---|
| | eV | | lux | V | $\mu Acm^{-2}$ | | % | |
| PB2:ITCC | 1.72 | 3000K LED | 200 | 0.895 | 20.9 | 0.728 | 21.7 | [81] |
| | | | 500 | 0.919 | 51.8 | 0.756 | 23.2 | |
| | | | 1000 | 0.951 | 109.3 | 0.766 | 25.4 | |
| PB2:FTCC-Br | 1.72 | 3000K LED | 200 | 0.884 | 23.7 | 0.767 | 26.1 | |
| | | | 500 | 0.91 | 58.6 | 0.795 | 28.3 | |
| | | | 1000 | 0.943 | 118.6 | 0.811 | 30.2 | |
| PBDB-TF:PC71BM | 1.87 | 2700K LED | 200 | 0.712 | 18.9 | 0.713 | 15.9 | [82] |
| | | | 500 | 0.758 | 47.2 | 0.727 | 17.2 | |
| | | | 1000 | 0.784 | 94.1 | 0.741 | 18.1 | |
| PBDB-TF:IT-4F | 1.57 | 2700K LED | 200 | 0.659 | 22.8 | 0.734 | 18.2 | |
| | | | 500 | 0.692 | 56.6 | 0.756 | 19.6 | |
| | | | 1000 | 0.712 | 113 | 0.78 | 20.8 | |
| PBDB-TF:ITCC | 1.75 | 2700K LED | 200 | 0.918 | 19.2 | 0.7 | 20.4 | |
| | | | 500 | 0.948 | 47.8 | 0.706 | 21.2 | |
| | | | 1000 | 0.962 | 95.8 | 0.722 | 22 | |
| PBDB-TF:IO-4Cl | 1.80 | 2700K LED | 200 | 1.03 | 18.2 | 0.715 | 22.2 | [83] |
| | | | 500 | 1.07 | 45.1 | 44.7 | 24.6 | |
| | | | 1000 | 1.1 | 90.6 | 89.4 | 26.1 | |
| PBDB-TF:IT-4F | 1.57 | 2700K LED | 1000 | 0.712 | 114 | 0.789 | 21.2 | |
| PM6:Y6-O | 1.52 | 3000K LED | 290 | 0.79 | 44 | 0.71 | 28.1 | [84] |
| | | | 700 | 0.81 | 102 | 0.76 | 29.5 | |
| | | | 1200 | 0.83 | 175 | 0.76 | 30 | |
| | | | 1650 | 0.84 | 245 | 0.76 | 30.9 | |
| P3TEA: FTTB-PDI4 | 1.66 | 3000K LED | 290 | 0.95 | 32 | 0.65 | 22.5 | |
| | | | 700 | 0.99 | 79 | 0.67 | 24.7 | |
| | | | 1200 | 1.02 | 143 | 0.67 | 26.2 | |
| | | | 1650 | 1.02 | 196 | 0.67 | 26.7 | |
| PPDT2FBT:PCBM | 1.80 | CCT 5600K LED | 300 | 0.58 | 26.8 | 0.672 | 11.5 | [85] |
| | | | 600 | 0.61 | 52.3 | 0.69 | 12.1 | |
| | | | 1000 | 0.62 | 85 | 0.695 | 11.8 | |
| | | | 3000 | 0.65 | 307 | 0.696 | 14.9 | |
| | | | 5000 | 0.68 | 404 | 0.72 | 12.7 | |
| | | | 10000 | 0.71 | 835 | 0.722 | 13.7 | |
| PPDT2FBT:ITIC-M | 1.57 | CCT 5600K LED | 300 | 0.53 | 20.8 | 0.57 | 6.9 | |
| | | | 600 | 0.58 | 42.4 | 0.566 | 7.6 | |
| | | | 1000 | 0.62 | 68.5 | 0.546 | 7.5 | |
| | | | 3000 | 0.7 | 227 | 0.479 | 8.2 | |
| | | | 5000 | 0.74 | 320 | 0.53 | 8.1 | |
| | | | 10000 | 0.79 | 649 | 0.528 | 8.7 | |
| PPDT2FBT:ITIC-F | 1.52 | CCT 5600K LED | 300 | 0.29 | 34.8 | 0.313 | 3.5 | |
| | | | 600 | 0.36 | 51.6 | 0.338 | 3.5 | |
| | | | 1000 | 0.45 | 85.5 | 0.376 | 4.7 | |
| | | | 3000 | 0.57 | 297 | 0.488 | 8.9 | |
| | | | 5000 | 0.63 | 404 | 0.506 | 8.3 | |
| | | | 10000 | 0.67 | 817 | 0.555 | 9.8 | |
| PPDT2FBT:tPDI2N-EHa | 1.80 | CCT 5600K LED | 300 | 0.79 | 20.9 | 0.499 | 9 | |
| | | | 600 | 0.82 | 40.3 | 0.606 | 9.2 | |
| | | | 1000 | 0.84 | 65.4 | 0.502 | 8.9 | |
| | | | 3000 | 0.88 | 187 | 0.504 | 9 | |
| | | | 5000 | 0.9 | 316 | 0.498 | 9.1 | |
| | | | 10000 | 0.93 | 650 | 0.488 | 9.5 | |
| PPDT2FBT:PCBM | 1.80 | CCT 2800K LED | 300 | 0.59 | 29.4 | 0.675 | 13 | |
| | | | 600 | 0.61 | 57.7 | 0.692 | 12.5 | |
| | | | 1000 | 0.62 | 94.6 | 0.698 | 13.8 | |
| | | | 3000 | 0.67 | 272 | 0.718 | 14.7 | |
| | | | 5000 | 0.69 | 456 | 0.721 | 15.2 | |
| | | | 10000 | 0.71 | 968 | 0.724 | 16.6 | |
| PPDT2FBT:ITIC-M | 1.57 | CCT 2800K LED | 300 | 0.54 | 24.1 | 0.57 | 8.5 | |
| | | | 600 | 0.59 | 46.8 | 0.557 | 8.6 | |
| | | | 1000 | 0.62 | 77.4 | 0.544 | 8.9 | |
| | | | 3000 | 0.72 | 216 | 0.538 | 9.4 | |
| | | | 5000 | 0.76 | 364 | 0.531 | 9.9 | |
| | | | 10000 | 0.8 | 765 | 0.531 | 10.8 | |
| PPDT2FBT:ITIC-F | 1.52 | CCT 2800K LED | 300 | 0.26 | 30.6 | 0.303 | 2.7 | |
| | | | 600 | 0.37 | 59.1 | 0.338 | 4.1 | |
| | | | 1000 | 0.45 | 96.9 | 0.372 | 5.4 | |
| | | | 3000 | 0.59 | 276 | 0.448 | 8.2 | |
| | | | 5000 | 0.63 | 456 | 0.489 | 9.5 | |
| | | | 10000 | 0.66 | 975 | 0.535 | 11.6 | |
| PPDT2FBT:tPDI2N-EHa | 1.80 | CCT 2800K LED | 300 | 0.79 | 20.9 | 0.501 | 9.3 | |
| | | | 600 | 0.81 | 39.7 | 0.511 | 9.2 | |
| | | | 1000 | 0.84 | 66.8 | 0.51 | 9.6 | |
| | | | 3000 | 0.88 | 219 | 0.442 | 9.6 | |
| | | | 5000 | 0.9 | 308 | 0.506 | 9.5 | |
| | | | 10000 | 0.93 | 664 | 0.496 | 10.2 | |
| PTB7-Th:PCBM | 1.58 | LED | 186 | 0.56 | 19 | 0.72 | 10.6 | [86] |
| | | | 890 | 0.62 | 92 | 0.74 | 11.6 | |
| c-Si | 1.12 | LED | 186 | 0.37 | 22 | 0.63 | 6.92 | |
| | | | 890 | 0.43 | 120 | 0.71 | 9.65 | |
| PM6:Y6-O | 1.52 | LED | 250 | 0.825 | 32.2 | 0.8 | 29.2 | [87] |
| | | | 500 | 0.846 | 64.3 | 0.811 | 30.3 | |
| | | | 1000 | 0.866 | 128.6 | 0.815 | 31.2 | |
| PM6:IT-4F | 1.57 | 3000K LED | 250 | 0.65 | 36.74 | 0.6115 | 20.2 | [88] |
| | | | 500 | 0.68 | 733.33 | 0.6424 | 23.4 | |
| | | | 1000 | 0.71 | 138.72 | 0.6873 | 23.75 | |
| | | 6000K LED | 1000 | 0.69 | 126.89 | 0.6724 | 19.73 | |

## Appendix – Dark Saturation Current Density in Organic Photovoltaics

In the weak absorption limit, we take the spectral lineshape of the sub-gap photovoltaic external quantum efficiency to be given by Equation **(S20)**. In the non-generate limit, the corresponding dark saturation current density in the radiative limit will be given by

$$J_0^{\mathrm{rad}} \approx \frac{\mathrm{EQE_{max}}\pi q}{h^3c^2}\int_0^{\infty} E^2 \exp\left(-\frac{E}{kT}\right)\left[\exp\left(\frac{E-E_{\mathrm{opt}}+\frac{\sigma_s^2}{2kT}}{kT}\right)\mathrm{erfc}\left(\frac{E-E_{\mathrm{opt}}+\frac{\sigma_s^2}{kT}}{\sigma_s\sqrt{2}}\right) + \mathrm{erf}\left(\frac{E_{\mathrm{opt}}}{\sigma_s\sqrt{2}}\right) + \mathrm{erf}\left(\frac{E-E_{\mathrm{opt}}}{\sigma_s\sqrt{2}}\right)\right]\mathrm{d}E. \tag{A1}$$

This integral can be evaluated in three parts, the simplest of which is evaluated using the gamma function:

$$\int_0^{\infty} E^2 \exp\left(-\frac{E}{kT}\right)\mathrm{d}E = 2(kT)^3. \tag{A2}$$

The next integral may be evaluated by-parts, where the derivative of the error function needs to be taken. With $x = \frac{E}{\sigma_s\sqrt{2}}$ and $\Delta = -E_{\mathrm{opt}} + \frac{\sigma_s^2}{kT}$, this gives

$$\begin{aligned}\int_0^{\infty} E^2\mathrm{erfc}\left(\frac{E-E_{\mathrm{opt}}+\frac{\sigma_s^2}{kT}}{\sigma_s\sqrt{2}}\right)\mathrm{d}E &= \frac{2\left(\sigma_s\sqrt{2}\right)^3}{3\sqrt{\pi}}\int_0^{\infty} x^3 \exp\left(-\left[x+\frac{\Delta}{\sigma_s\sqrt{2}}\right]^2\right)\mathrm{d}x \\ &= \frac{\left(\sigma_s\sqrt{2}\right)^3}{3}\left(\frac{1}{\sqrt{\pi}}\left[1+\frac{\Delta^2}{2\sigma_s^2}\right]\exp\left(-\frac{\Delta^2}{2\sigma_s^2}\right) - \frac{\Delta}{\sigma_s\sqrt{2}}\left[\frac{3}{2}+\frac{\Delta^2}{2\sigma_s^2}\right]\mathrm{erfc}\left(\frac{\Delta}{\sigma_s\sqrt{2}}\right)\right).\end{aligned} \tag{A3}$$

To obtain the final line, the following integrals were used:

$$\int e^{-x^2}\,\mathrm{d}x = \frac{\sqrt{\pi}}{2}\mathrm{erf}(x) + C, \tag{A4a}$$

$$\int xe^{-x^2}\,\mathrm{d}x = -\frac{e^{-x^2}}{2} + C, \tag{A4b}$$

$$\int x^n e^{-x^2}\,\mathrm{d}x = -\frac{x^{n-1}e^{-x^2}}{2} + \frac{(n-1)}{2}\int x^{n-2}e^{-x^2}\mathrm{d}x, \quad \text{for} \quad n \geq 2 \tag{A4c}$$

The final of the three integrals is the most involved. With some variable substitutions $x = \frac{E}{kT}$, $\delta = -\frac{E_{\text{opt}}}{\sigma_{\text{s}}\sqrt{2}}$, and $r = \frac{kT}{\sigma_{\text{s}}\sqrt{2}}$, integration by parts gives

$$\int_0^\infty E^2 \,\text{erf}\left(\frac{E - E_{\text{opt}}}{\sigma_{\text{s}}\sqrt{2}}\right) \exp\left(-\frac{E}{kT}\right) \text{d}E$$

$$= (kT)^3 \Bigg( 2\,\text{erf}(\delta) + \frac{2r}{\sqrt{\pi}} \exp\left(\frac{\Delta^2 - E_{\text{opt}}^2}{2\sigma_{\text{s}}^2}\right) \int_0^\infty [x^2 + 2x + 2] \exp\left(-\left[rx + \frac{\Delta}{\sigma_{\text{s}}\sqrt{2}}\right]^2\right) \text{d}x \Bigg). \quad \textbf{(A5)}$$

The integrals can be evaluated as above. After combining with the result of the first integral, making use of the fact that the error function is an odd function, and simplifying, one finds

$$\text{erf}\left(\frac{E_{\text{opt}}}{\sigma_{\text{s}}\sqrt{2}}\right) \int_0^\infty E^2 \exp\left(-\frac{E}{kT}\right) \text{d}E + \int_0^\infty E^2 \,\text{erf}\left(\frac{E - E_{\text{opt}}}{\sigma_{\text{s}}\sqrt{2}}\right) \exp\left(-\frac{E}{kT}\right) \text{d}E$$

$$= \exp\left(\frac{\Delta^2 - E_{\text{opt}}^2}{2\sigma_{\text{s}}^2}\right) \Bigg( \sigma_{\text{s}} \sqrt{\frac{2}{\pi}} [-\Delta kT + 2(kT)^2] \exp\left(\frac{\Delta^2}{2\sigma_{\text{s}}^2}\right) + \left[kT\left(\Delta^2 + \sigma_{\text{s}}^2\right) - 2\Delta(kT)^2 + 2(kT)^3\right] \text{erfc}\left(\frac{\Delta}{\sigma_{\text{s}}\sqrt{2}}\right) \Bigg). \quad \textbf{(A6)}$$

Finally, combining all the pieces gives the full expression for the dark saturation current density,

$$J_0^{\text{rad}} \approx \frac{\text{EQE}_{\max}\pi q}{h^3 c^2} \exp\left(\frac{-E_{\text{opt}} + \frac{\sigma_{\text{s}}^2}{2kT}}{kT}\right) \Bigg[ \sigma_{\text{s}} \sqrt{\frac{2}{\pi}} \left(\frac{\Delta^2}{3} - \Delta kT + 2k^2T^2 + \frac{2\sigma_{\text{s}}^2}{3}\right) \cdot \exp\left(-\frac{\Delta^2}{2\sigma_{\text{s}}^2}\right) + \left((\Delta^2 + \sigma_{\text{s}}^2)kT - \Delta[2k^2T^2 + \sigma_{\text{s}}^2] + 2k^3T^3 - \frac{\Delta^3}{3}\right) \text{erfc}\left(\frac{\Delta}{\sigma_{\text{s}}\sqrt{2}}\right) \Bigg]. \quad \textbf{(A7)}$$